\documentclass[preprint]{aastex}
\begin{document}

\received{8 October 2000}
\accepted{27 November 2000}
\slugcomment{To appear in the Astronomical Journal}

\def\gsim{\;\lower.6ex\hbox{$\sim$}\kern-7.75pt\raise.65ex\hbox{$>$}\;
}
\def\lsim{\;\lower.6ex\hbox{$\sim$}\kern-7.75pt\raise.65ex\hbox{$<$}\;
}
\def\aa{A\&A }
\def\pasj{PASJ}
\def\mb{$m_{\rm{F439W}}$}
\def\mv{$m_{\rm{F555W}}$}
\def\mj{$m_{\rm{F110W}}$}
\def\mh{$m_{\rm{F160W}}$}
\def\mbv{$m_{\rm{F439W}}-m_{\rm{F555W}}$}
\newcommand{\MSUN}{M$_{\odot}$}
\newcommand{\ZSUN}{Z$_{\odot}$}
\newcommand{\Myr}{M$_{\odot}$~yr$^{-1}$}
\newcommand{\Myk}{M$_{\odot}$~yr$^{-1}$~kpc$^{-2}$}

\title{The red stellar population in NGC~1569
\footnote{Based on observations with the NASA/ESA Hubble
Space Telescope, obtained at the Space Telescope Science Institute,
which is operated by AURA for NASA under contract NAS5-26555}}

\vspace{0.5in}

\author{A. Aloisi$^{1}$, M. Clampin$^1$, E. Diolaiti$^2$, L.
Greggio$^{3,4}$, Claus Leitherer$^1$, A. Nota$^{1,5}$, L. Origlia$^3$, 
G. Parmeggiani$^3$, and M. Tosi$^3$}

\vspace{0.25in}

\affil{$^1$ Space Telescope Science Institute,
       3700 San Martin Drive, Baltimore, MD 21218\\
       e-mail: aloisi@stsci.edu, clampin@stsci.edu,
       leitherer@stsci.edu, nota@stsci.edu}

\affil{$^2$ Dipartimento di Astronomia, Universit\`a di Bologna, Via
       Ranzani 1, I-40127 Bologna, Italy\\
       e-mail: diolaiti@bo.astro.it}

\affil{$^3$ Osservatorio Astronomico di Bologna, Via Ranzani 1,
       I-40127 Bologna, Italy\\
       e-mail: greggio@bo.astro.it, origlia@bo.astro.it,
               parmeggiani@bo.astro.it, tosi@bo.astro.it}

\affil{$^4$ Universitaets Sternwarte Muenchen,
       Scheinerstrasse 1, D-81679 Muenchen, Germany\\
       e-mail: greggio@usm.uni-muenchen.de}

\affil{$^5$ Affiliated with the Astrophysics Division, Space Science 
Department of the European Space Agency}

\begin{abstract}

We present HST NICMOS photometry of the resolved stellar population in 
the dwarf irregular galaxy NGC~1569. The color-magnitude diagram (CMD)
in the F110W and F160W photometric bands contains $\sim$2400 stars with 
a formal photometric error $\lsim\,0.1$~mag down to \mj~$\approx 23.5$ and 
\mh~$\approx 22.5$. The fiducial photometry has a completeness factor 
higher than 50\% down to \mj~$\approx 21.5$ and \mh~$\approx 20.0$. 
We describe the data processing which is required to calibrate the 
instrumental peculiarities of NICMOS. Two different packages (DAOPHOT
and StarFinder) for PSF-fitting photometry are used to strengthen the 
photometric results in the crowded stellar field of NGC~1569. 

The resulting CMD is discussed in terms of the major evolutionary properties 
of the resolved stellar populations. For a distance modulus of $(m-M)_0=26.71$ 
and  a reddening of $E(B-V)=0.56$, our CMD samples stars down to $\sim$\,0.8 
\MSUN, corresponding to look-back times of more than 15~Gyr (i.e., an entire 
Hubble time). This is clear indication of star-formation activity in NGC~1569 
spanning an entire Hubble time. The metallicity of the reddest red giant branch 
(RGB) stars is in better agreement with $Z=0.004$ as measured in HII regions, 
than with  $Z=0.0004$ as expected from the stellar ages. The presence of --- 
yet undetected --- very metal-poor stars embedded in the stellar distribution 
around \mj~=~22.75 and \mj~--~\mh~=~1.15 is, however, not ruled out. The youngest 
stars ($\lsim$\,50~Myr) are preferentially found around the two central super star 
clusters, whereas the oldest population has a more uniform spatial distribution. 
A star-formation rate per unit area of 1~M$_{\odot}$\,yr$^{-1}$\,kpc$^{-2}$ and a 
mass formed in stars of $\sim1.4 \times 10^6$~\MSUN~ in the last 50~Myr are derived 
from the CMD. The near-infrared (NIR) CMD places strong constraints on the lower 
limit of the onset of star formation in NGC~1569.

The exceptionally high crowding in the NICMOS images of NGC~1569 is a challenge 
for the photometric analysis. As a result, optical and NIR images of NGC~1569 
sample different populations and cannot be cross-correlated. Nevertheless, 
we demonstrate the consistency of the star-formation histories derived from the 
optical and NIR CMDs.

\end{abstract}


\section{Introduction}

Dwarf irregular and blue compact galaxies are important for our understanding of 
galaxy formation and evolution. In the models of hierarchical clustering (Kauffmann 
et al. 1993; Cole et al. 1994), low-mass dwarfs are the first structures to collapse 
from small-scale primordial density fluctuations in the early universe, giving birth 
to larger stellar systems through merging. The low metallicity (Kunth 1985; Thuan et 
al. 1994) and high gas content (Thuan \& Martin 1981) of nearby dwarf star-forming 
systems suggests a rather low level of their chemical evolution, and makes some of 
them (e.g., I~Zw~18 or SBS 0335\,--\,052) plausible counterparts to primeval galaxies in 
the early universe (Izotov \& Thuan 1999). Starbursting systems could represent the 
local examples of the faint blue galaxies found in deep imaging surveys as an excess 
with respect to standard models (e.g., Tyson 1988; Babul \& Ferguson 1996; Williams 
et al. 1996). Recent analyses have shown that the excess counts are possibly due to 
the intrinsically bluest, most compact systems, undergoing strong bursts of star 
formation (SF) at redshift $z\gsim0.1$ (e.g., Lilly et al. 1995). Therefore, 
understanding how the SF proceeds in blue compacts and irregulars (see Tosi 1999 
for a review on the different SF regimes) is not only important in its own right but 
has general implications for astrophysical and cosmological issues.

The presence or absence of an underlying older population in dwarf systems with current 
SF is crucial for our understanding of galaxy evolution. Did all dwarf galaxies form 
stars already 10~--~15~Gyr ago and have they experienced discontinuous and/or low-level 
SF activity since then (Thuan 1983; Legrand 2000)? Have the metal-poorest systems just 
recently ($\lsim$\,100 Myr ago) started forming stars (Searle \& Sargent 1972; Searle, 
Sargent, \& Bagnuolo 1973; Izotov \& Thuan 1999)? If an old stellar population is present 
in all dwarf star-forming systems, we would have an indirect validation of the hierarchical 
collapse model. Identifying local dwarf starbursting galaxies as counterparts to the excess 
faint blue galaxy population at intermediate redshift would be more straightforward.
However, the SF rate (SFR) would have to be at least 1~\MSUN~yr$^{-1}$ at z$\approx$1 in 
order to make the brightness consistent with that of the intermediate redshift population
(Babul \& Ferguson 1996). Primordial systems in our local universe could possibly be
identified with HI high-velocity clouds (HVCs). These have been suggested by Blitz et al. 
(1999) to be just gas contained within dark-matter mini-halos, moving along filaments and 
accreting onto the Local Group. A physical mechanism would instead be required for dwarf 
systems to justify the onset of the SF regime only in relatively recent epochs (i.e. Babul 
\& Rees 1992). These ``young'' galaxies clearly could not contribute to samples at higher 
redshift ($z\gsim0.1$). 

The best way to search for old stellar populations is to observe the parent galaxies 
in the infrared (IR), where the reddening effects are much less severe than in the
optical, and to derive the corresponding CMDs, where the old low-mass stars in the RGB 
evolutionary phase are more easily visible and distinguishable from the younger, more 
massive objects. The brightest feature of old stellar populations on the CMD is the tip 
of the RGB (TRGB), which is located at a constant luminosity of $\log L/L_\odot\simeq 3.4$ 
for ages beyond $\simeq$\,1~Gyr (Sweigart, Greggio, \& Renzini 1990). All the SF prior to 
this epoch contributes stars at the TRGB, producing a jump in the LF of the red stars. 
Thus, the detection of the TRGB unambiguously shows that the parent galaxy harbors stellar 
populations older than 1~Gyr. The non-detection could imply only recent SF (i.e., ages 
younger than 1~--~2~Gyr), although the TRGB feature may be difficult to detect in the 
presence of a strong intermediate-age population.

In this paper we report on our NIR observations of the resolved stellar population 
in the core of the dwarf starbursting galaxy NGC~1569 (= UGC~3056 = Arp~210 = 
VII~Zw~16 = IRAS~4260\,+\,6444). This object was highlighted by Gallagher, Hunter, 
\& Tutukov (1984) as an outstanding object in their sample of actively star-forming
galaxies. It is a prime candidate for a detailed study of the SF history in a starburst 
galaxy, since at $D=(2.2\pm0.6)$~Mpc it is the closest starburst system outside the 
Local Group (Israel 1988). Some of the properties of NGC~1569 are typical of a dwarf 
galaxy. With a distance modulus of $(m-M)_0 = 26.71$, its total absolute $B$ magnitude 
is $M_{\rm{B,0}}\approx -17$ (Israel 1988), which is intermediate between the two 
Magellanic Clouds. Total mass, hydrogen content, and metallicity are estimated to be 
$M \approx 3.3 \times 10^8$~\MSUN, $M_{\rm H} \approx 1.3 \times 10^8$~\MSUN~ (Israel 
1988), and $Z \approx 0.25$~Z$_{\odot}$, respectively (Calzetti, Kinney, \& 
Storchi-Bergmann 1994; Gonz\'alez-Delgado et al. 1997). Thus, NGC~1569 is a gas rich 
system in a relatively early stage of chemical evolution.

NGC~1569 hosts the two ``super star clusters'' (SSCs) NGC~1569-A (consisting of the two 
sub-clusters NGC~1569-A1 and NGC~1569-A2) and NGC~1569-B. These high-density stellar 
clusters were first detected and discussed by Arp \& Sandage (1985) and Prada, Greve, 
\& McKeith (1994) and, more recently, by Gonz\'alez-Delgado et al. (1997). HST studies 
were done by O'Connell, Gallagher, \& Hunter (1994), Ho \& Filippenko (1996), De~Marchi 
et al. (1997), Buckalew et al. (2000), Hunter et al. (2000), and Origlia et al. (2000). 
The SSCs reflect a recent starburst in which at least $10^5$~\MSUN~ of gas was transformed 
into stars within about 1~pc. SSCs are similar to young Galactic globular clusters observed 
shortly after their formation (Melnick, Moles, \& Terlevich 1985; Meurer 1995; but see 
Zhang \& Fall 1999).

The recent strong SF activity around the SSCs of NGC~1569 is also found in multiwavelength 
studies of the different phases of the  interstellar medium (ISM). The cool gaseous phase 
is anticorrelated with the recent SF. The HI distribution shows a hole with a diameter of 
250~pc, centered on SSC A and encompassing SSC B (Israel \& van Driel 1990; Stil \& Israel 
1998). The HI, as well as the CO, is preferentially found in the proximity of the most 
extended HII region of NGC~1569, 90~pc to the west of SSC A (Young, Gallagher, \& Hunter 
1984; Greve et al. 1996; Taylor et al. 1999). Deep H$\alpha$ images indicate that the warm 
ionized gas is structured into filaments and arcs with lengths of several kiloparsecs, oriented 
along the minor axis of the galaxy (Hunter, Hawley, \& Gallagher 1993; Devost et al. 1997). 
Typical velocities around 50~--~100~km~s$^{-1}$ and maximum values as high as 200~km~s$^{-1}$ 
have been attributed to expanding superbubbles with dynamical ages of 10~Myr (Tomita, Ohta, \& 
Saito 1994; Heckman et al. 1995). Observations of the hot gas (10$^7$~K) with ROSAT and ASCA 
(Heckman et al. 1995; Della Ceca et al. 1996)  show that $\sim$~60\% of the soft X-ray 
emission in NGC~1569 is centered in between the two SSCs. Its origin is attributed to thermal 
emission from the hot gas of a galactic superwind emanating from the disk. The X-ray emission 
as well as the H$\alpha$ shells are located in the region where  HI and CO  are absent, another 
indication of the association of the hot and warm gas  with a galactic outflow. 

Observations of the  field stellar population of NGC~1569 are required to trace the
SF activity over look-back times greater than 10~Myr. The field stellar population (i.e., 
outside the unresolved SSCs and HII regions) has been resolved with HST by Vallenari \& 
Bomans (1996, VB) on WFPC $V$ and $I$ images, and by Greggio et al. (1998, G98) on WFPC2 
$B$ and $V$ images. VB demonstrated that the galaxy experienced a strong starburst from 
0.1~Gyr to 4~Myr ago by comparing isochrones to the derived CMD. The well populated asymptotic 
giant branch (AGB) definitely argues for previous SF in NGC~1569. The authors conclude that a 
prior SF episode  between 1.5 and 0.15~Gyr ago occurred at a significantly lower rate than 
the most recent starburst. However, the values of these parameters are uncertain due to 
observational limitations.  G98 interpreted the data with the synthetic CMD method and showed 
that in the last 0.15~Gyr the field has experienced a general SF activity of about 0.1~Gyr
duration and roughly constant intensity. The SF activity stopped approximately 5~--~10~Myr 
ago. The estimated SFR during this episode is 4~--~20~\Myr (depending on the slope of the 
initial mass function, IMF), or 0.5~--~3~\Myr~kpc$^{-2}$ for an area of 0.14~kpc$^2$ and a 
distance of 2.2~Mpc (G98). This value is about 1000 times higher than the current average 
SFR in our Galaxy and at least two orders of magnitude higher than in Local Group irregulars. 
A close low-density HI-cloud companion, which is connected to NGC~1569 by an HI bridge, may 
have triggered the exceptionally high SF activity by interaction (Stil \& Israel 1998).

The SF history for epochs older than 0.1~Gyr can be studied by searching for AGB and 
RGB stars in the IR. We thus observed this exceptional starbursting dwarf galaxy with 
the NICMOS camera on board of HST. The NICMOS field of view was centered on our previous 
PC field. This paper presents the results of this study. Section~2 describes the data 
collection and reduction. The photometry applied to the two reduced images in the F110W 
and F160W filters is in Section~3. Two different PSF-fitting photometry packages were 
used to assess the reliability of the derived magnitudes. The results of the completeness 
tests are described. Section~4 contains an analysis of the NIR CMDs and luminosity function 
(LF). Comparison of these data with stellar evolutionary tracks allows an estimate of the 
stellar masses, time scales, and metallicities involved in the SF process of this galaxy. 
Section~5 discusses the cross-correlation between the NIR and optical data of G98. In 
Section~6 the properties of the resolved stellar population are correlated with the overall  
starburst activity. Section~7 is a discussion  of all the major factors contributing to 
making NGC~1569 a challenging case for the photometric reduction. A summary of the major 
results is in Section~8.

\section{NICMOS observations and data reduction}

The observations were performed on 25 February, 1998 using the NICMOS/NIC2 camera on 
board of HST. NIC2 has a field of view of 19$\farcs$2 $\times$ 19$\farcs$2, and a pixel 
size of 0$\farcs$075. The telescope was pointed to place all the NGC~1569 SSCs in the NIC2 
field of view. The field of view covers the central part of the WFPC2/PC images which were 
previously obtained by G98. The observations were made using the NICMOS broad band filters 
F110W and F160W. A description of the NICMOS cameras and their corresponding filter sets is 
given by Calzetti et al. (1999).

Ten exposures were obtained with a total on-source integration time of 85~min in each filter. 
Each exposure was taken with the MULTIACCUM readout and consists of 13 non-destructive reads 
of the array from 0.0 seconds (the bias level) to 512 seconds using the pre-defined sequence 
STEP256. A spiral dither pattern of 0$\farcs$2 was adopted to account for pixel to pixel 
non-uniformities. The exposures were planned using STScI's NICMOS Exposure Time Calculator to 
obtain stellar photometry in the $J$ and $H$ bands down to 23~--~24~mag, with a S/N~$>$~10. For 
a distance of $D = 2.2$~Mpc, these exposures reach the tip of the RGB, which is expected at 
$\log(L/L_{\odot})= 3.4$, with a margin of two magnitudes.

The initial analysis was performed on the mosaicked images from the automatic CALNICA and 
CALNICB pipelines, which are part of the STSDAS package in IRAF\footnote{IRAF is distributed 
by the National Optical Astronomy Observatories, which are operated by AURA, Inc., under 
cooperative agreement with the National Science Foundation.}. These are standard reduction 
procedures which remove the instrumental signature and co-add datasets obtained from multiple 
iterations of the same exposures (CALNICA), and  mosaicked images  obtained from dither 
patterns (CALNICB). CALNICA performs bias and dark subtraction, non-linearity and flat-field 
corrections, count rate conversion, and photometric calibration. In the case of MULTIACCUM 
observations, it performs also cosmic-ray identification and rejection, and saturation 
correction. CALNICB estimates and removes the background, determines the offsets between 
dithered images, and aligns them by performing a  bilinear interpolation. 

In addition, the  processed images are affected by a number of  instrumental peculiarities
which are not corrected for by CALNICA. While the automatic pipelines take care of all 
instrumental effects which are stable in time, they do not correct for those which are 
known to vary. Among these are  {\it shading} and variable bias, also called {\it pedestal}.

$\bullet$ {\it Shading} is a noiseless signal gradient, like a pixel-dependent bias, which 
changes in the direction of the pixel clocking during a readout. This bias is imparted by 
the readout amplifiers and is temperature dependent. Its amplitude is large in NIC2 (up to 
several hundred electrons across a quadrant). Throughout the instrument lifetime, the 
temperature of NICMOS has varied by a couple of degrees. CALNICA assumes a dark frame for 
the bias\,+\,dark subtraction which includes a shading profile at 61.4~K (October 1997).
Observations taken significantly later, such as the data discussed here, can display a 
residual shading due to a partial correction resulting from a shading profile for a slightly 
different temperature (Calzetti et al. 1999). In order to solve this problem, a new shading 
profile was generated which was appropriate for the temperature at the time of the observation. 
Then, the CALNICA module responsible for bias and dark subtraction was reapplied.

$\bullet$ The {\it pedestal} is a variable additive offset introduced during array reset, 
in addition to the net quadrant bias. It is believed that these changes may be thermally 
driven and are either in the electronics or the detectors themselves or a combination of 
both. In the case of a MULTIACCUM sequence such as the one obtained for the observations 
of NGC~1569, the net bias change over the course of the exposure results in an additive 
offset to the final image. When CALNICA flat fields the images of each single readout, 
the bias offset in each image is modulated by the flat field and appears as an inverse 
flat-field pattern in the calibrated final image. The pedestal can be easily removed in 
situations where the bias offset can be determined in image regions with few stars. In 
extremely crowded fields, such as NGC~1569, its removal is more problematic. We have 
applied the Pedestal Estimation and Quadrant Equalization Software developed by Roeland 
P. van der Marel for crowded fields 
(see details at the URL http://sol.stsci.edu/~marel/software/pedestal.html). 
This software considers a flat-fielded NICMOS image, and estimates the amount of 
any constant bias level (the pedestal) that may be still present in the data, by 
minimizing the amount of falsely imprinted flat-field pattern on both large and 
pixel-to-pixel scales with an iterative process. The best-fitting constant, divided
by the flat field, is then subtracted from the calibrated image. An algorithm that 
attempts to remove differences between quadrants is also included. In the case of 
NGC~1569 the average over the whole field of the variable bias taken away from each 
single exposure is of the order of $\sim$~10\% in both filters, with peaks as high 
as $\sim$~45\% in the regions mostly devoid of stars.

The presence of additional effects, such as cosmic-ray persistence and non-linearity
was also assessed.

$\bullet$ {\it Cosmic ray persistence:} During regular passages through the South 
Atlantic Anomaly (SAA), the NICMOS detectors are bombarded with cosmic rays which 
deposit a large signal into nearly every pixel.  The persistent signal from the 
cosmic rays may be present as a residual pattern in exposures taken less than 30 
minutes after a SAA passage. In the case of the NICMOS NGC~1569 observations, only 
four out of the twenty datasets were taken within 30 minutes after a SAA passage, 
and for all those  four cases at least 20 minutes had already elapsed. We therefore 
concluded that cosmic ray persistence does not affect the dataset, given the steep 
exponential decay of the CR persistence effect (Calzetti et al. 1999).

$\bullet$ {\it Linearity}: An improved correction for non-linearity has recently
been made available in CALNICA. This correction was not included in our original 
data reduction. A comparison with the old correction showed a marginal improvement 
of $\simeq$~2\% for the brightest stars only. Therefore the new correction has not 
been included.

NICMOS data of NGC~1569 are affected by severe crowding. This is evident by looking 
at Figure~\ref{imageH}, where we present the mosaicked image in F160W resulting from 
our final data reduction. The NIC2 PSF has a typical FWHM of $\simeq$~1.4, 1.8 pixels 
in the F110W, F160W filters, respectively. The corresponding angular sizes are 0$\farcs$11 
and 0$\farcs$14, respectively. Under extreme crowding conditions, such a sampling often 
prevents the standard photometry packages from detecting the faintest stars, since sometimes 
only the PSF peaks are visible. In order to detect all the stars down to the predicted 
limiting magnitude, it was necessary to take advantage of our dithered observations to 
improve the sampling of the PSF. The observations of NGC~1569 were taken by dithering the 
camera with a spiral pattern of step comparable to the PSF FWHM. For each filter, the 10 
dithered images, individually corrected for the effects listed above, were then combined 
to obtain a fully sampled image, using the software package {\it Drizzle} (Fruchter \& Hook 
1998). The PSF in the resulting {\it drizzled} images has a FWHM of 3.0 pixels in the F110W 
filter and 3.9 pixels in F160W (the effective pixel size is 0$\farcs$0375), and is thus 
better sampled. This makes it possible to fit and measure stellar magnitudes down to almost 
the faintest expected limit.

\section{The photometry}

\subsection{The photometric analysis}

The photometric reduction of the {\it drizzled} frames was performed using the IRAF
package DAOPHOT, and the same method followed by Aloisi, Tosi, \& Greggio (1999).
First, the automatic star detection routine DAOFIND was run on the final F110W
image, with the lowest possible detection  threshold (1$\,\sigma$ above the local
background). Although this selected threshold is lower than the values usually 
adopted ($\ga$\,\,3\,$\sigma$), only such a low threshold led to the retrieval of 
all objects visually identified in the frame. A higher initial threshold drastically 
reduced the number of detected objects. Therefore we lowered the threshold for the 
initial search and cleaned the sample subsequently: 5399 stars were retrieved in the 
combined F110W image by adopting the 1$\,\sigma$ threshold.

Photometry of all the stars found in the F110W image was then performed via PSF 
fitting. An initial estimate of the magnitude of each single star was obtained 
with an aperture photometry technique. We adopted a radius of 2 pixels. The pixel 
size of the {\it drizzled} image is 0$\farcs$0375. A template PSF was constructed 
by combining 20 isolated,  bright stars. The problem of neighboring stars was 
solved by using a $\sigma$-clipping algorithm to discard deviant single pixels in 
the calculation of the residuals from the analytic model. As a further check, we 
also produced an alternative template PSF by iteratively subtracting the neighboring 
stars, but found the resulting PSF profile to be unacceptably noisy due to the 
residuals in the star subtractions. The original template PSF was applied 
satisfactorily, and photometric measurements were obtained for a subset of 4375 
stars in the F110W filter.

The catalog with candidate stars to fit on the final {\it drizzled} image in the 
F160W filter was created from the positions of the stars previously fitted in F110W. 
These coordinates were used as input for the algorithms for the centering and the 
initial aperture photometry. The PSF fitting was performed in a similar fashion and 
photometric measurements were obtained for 3770 stars. After recentering, some mismatches 
were found in the centroids of the stars in the two filters: the two routines DAOMATCH 
and DAOMASTER were then employed to correctly match the coordinates of the stars in both 
filters. All stars showing an offset larger than the match radius (set to half the FWHM 
of the PSF, $\sim$\,1.5 pixels, or 0$\farcs$056) were discarded. After the match, 3492 
stars were retained. 

A reliable estimate of the local background for each star during the PSF-fitting 
photometry is important in crowded fields. In order to address this issue we 
calculated the sky contribution by considering two different techniques. In the 
standard procedure the local background was determined by taking an annulus with 
an inner radius rather close to the stellar radius (15 pixels), in order to avoid 
contamination from the stellar flux. An outer radius of 17 pixels was chosen in this 
case. In the {\it core PSF-fitting photometry} technique (De Marchi et al. 1993) the 
sky annulus was taken just beyond the first diffraction ring: between the 7th and 9th 
pixel from the central peak in F110W, and between the 9th and 11th pixel in F160W. 
An offset was then added to the calibrated magnitudes to take into account the 
fraction of source light present in the sky determination: --\,0.009 and --\,0.008 
in F110W and F160W, respectively. With the {\it core PSF-fitting photometry} about 
20\% (5\%) of the stars have magnitudes that differ by more than 0.1 (0.3) from those 
derived with the standard technique. This is a clear indication of the difficulties 
encountered in the correct estimate of the background. The distribution of the objects 
on the resulting CMDs from the two different background estimates, is, however, the same. 
Also the distribution of the formal errors for each filter is very similar. We have thus 
decided to maintain our first choice for the background evaluation.

A final screening of the F110W and F160W photometric measurements was done using the
$\chi^2$ and  the {\it sharpness} criteria. The $\chi^2$ index compares the observed 
pixel-to-pixel scatter in the fit residuals with the scatter calculated from a predictive 
model based on the measured detector characteristics.  In order to infer the appropriate 
$\chi^2$ values for our data set, we inspected the distributions of the $\chi^2$ values
in each filter as a function of magnitude (Figure~\ref{chi}). A conservative threshold 
$\chi^2\!<\,$1.5 applied to each filter data set resulted in eliminating the large 
majority of the spurious objects. Objects rejected by this criterion were typically the 
brightness peaks in the two SSCs, smaller star clusters, extended light distributions, or 
blends. 

In a similar fashion, we considered the {\it sharpness} criterion, which sets the intrinsic 
angular size of the astronomical object. Again, we plotted the {\it sharpness} parameter as 
a function of magnitude for each filter, and studied the resulting distributions, shown in 
Figure~\ref{sharpness}. A {\it sharpness} value between --\,0.4 and +\,0.4 in both filters 
naturally defines the boundary between the bulk of true stars and spurious objects. In this 
case, the excluded objects were typically found to be noise spikes ({\it sharpness}$\,<$~--\,0.4) 
and a few blends and peaks in clusters and SSCs ({\it sharpness}$\,>$~+\,0.4), erroneously 
identified as single stars. All the spurious detections can have some of the shape parameters 
with acceptable values at least in one filter, depending on what is contributing to them. 
For example the center of both SSCs has been rejected only because of the $\chi^2$ and 
{\it sharpness} values in F110W. The {\it sharpness} distribution in F160W is better than 
that in F110W. This can be due to the fact that the PSF in F160W has a larger FWHM, hence
is somewhat better sampled.

At the end of this screening, our catalog had 3177 stars with measured photometry in the F110W 
and F160W filters. In Figure~\ref{error} we show the formal photometric errors $\sigma_{\rm{DAO}}$ 
assigned by DAOPHOT. As our theoretical interpretation of the observed CMDs is more reliable if 
applied to stars with a small associated photometric error, we discarded  all objects with a DAOPHOT 
error larger than $0.1$~mag in either filter. With this selection, our final catalog contains 2423 
stars common to both filters, down to a limiting magnitude \mj~=\,23.5, and \mh~=\,22.5.

\subsection{The photometric calibration}

The instrumental magnitudes $m_{\rm i}$ from the PSF-fitting procedure are already normalized 
to the exposure time. They were converted into the calibrated magnitudes in the HST VEGAMAG 
system $m$ by applying the following formula (see Dickinson 1999 for more details):

\begin{center}
$m = m_{\rm i} + C_{\rm ap} + C_{\infty} + ZP_{\rm V}$.
\end{center}

$C_{\rm ap}$ is the aperture correction to convert our measurements from a 2 pixel radius 
in the {\it drizzled} images to the conventional radius of 0$\farcs$5 for NIC2 photometry.
$C_{\infty}$ is an  offset of --\,0.152 magnitudes added to translate the magnitude at a 
0$\farcs$5 radius into a nominal infinite aperture. This aperture is defined to encompass
1.15 times the flux measured in an aperture with 0$\farcs$5 radius. $ZP_{\rm V}$ is the 
published zero point for conversion into the HST VEGAMAG system: 22.381 and 21.750 for 
F110W and F160W, respectively.

The determination of the aperture correction is the most critical step in the photometric 
calibration of this crowded stellar field. To estimate its value, it is necessary to trace 
the encircled energy contained in well isolated stars as a function of distance from their 
center. For each filter we calculated an aperture correction directly from our PSF, 
considering it the best stellar template available. 

The NICMOS PSF is extremely stable, both in time and as a function of position with respect 
to the detector field of view. A quite reliable estimate of the aperture correction can be 
inferred from the encircled energy curves measured by Holfeltz \& Calzetti (1999) on a number 
of isolated, high S/N standard stars observed in the various filters as part of the NICMOS 
calibration program. We compared these published values with the aperture corrections calculated 
directly from the template PSF and found a discrepancy of 0.23 and 0.17~mag in the F110W and 
F160W filters, respectively. We ascribed the larger aperture correction values derived from the 
template PSF to contamination by neighboring stars. For this reason, we adopted the values of 
Holfeltz \&  Calzetti (1999) which were derived in optimal observing conditions.

The calibrated CMD in the HST F110W and F160W bands for the central region of NGC~1569
is presented in Figure~\ref{cmdnic}. The diagram in panel $a$ shows the 3177 objects
which were measured with DAOPHOT in both the F110W and F160W bands and were retained 
after the shape screening ($\chi^2<\,$1.5 and --\,0.4\,$<\,sharpness\,<$\,+\,0.4). The 
diagram in panel $b$ shows the subsample of 2423 objects with a small formal photometric 
error, $\sigma_{\rm{DAO}}<\,$0.1, in both filters.

\subsection{Photometry: confirmation by an alternative method}

The {\it drizzled} frames were  also analyzed with StarFinder, a new code 
developed for astrometry and photometry in crowded stellar fields. This code 
employs a different treatment of the effects due to stellar crowding and 
background contamination. It is described in detail in Diolaiti et al. (2000). 

With this code a numerical PSF template can be directly extracted by modelling the 
stars observed in the frame (similarly to DAOPHOT), or it can be simulated with Tiny 
Tim (Krist \& Hook 1999).  In the case of NGC~1569, the PSFs for the F110W and F160W 
filters were inferred from the images by computing the median average  of 10 suitable 
stars after correction for the local background and surrounding sources. A candidate 
object list was then created to include all brightness peaks at 3$\sigma$ above the 
background. The list was correlated with the template PSF in order of decreasing 
intensity. We accepted an object as a star if the correlation coefficient, which is a 
measure of the similarity with the PSF, is greater than a selected level between 0 and 
1. For this work, the correlation coefficient threshold was set to 0.7. This value was 
derived from an analysis of simulations as the best representation of the boundary between 
real stars and spurious detections.

During the PSF fitting, the code progressively creates a virtual copy of the observed 
field as a smooth background emission with superimposed stellar sources. The stars
are modeled as weighted shifted replicas of the PSF template and added one by one, 
in order of decreasing intensity. Photometry is performed on each individual star  
by using a sub-image of size comparable to the diameter of the first diffraction ring 
of the PSF. The local background is approximated by a bilinear surface, the underlying 
halos of brighter stars outside the fitting region are given by the synthetic image,  
and the brighter stars inside this region are represented as weighted shifted replicas 
of the PSF and re-fitted together with the analyzed star. The fainter candidate stars 
inside the fitting region are instead neglected at this point and are analyzed with the 
same strategy only in a further step. If the photometric fit is acceptable, the catalog 
and the synthetic image are updated with the new entry. For a better astrometric and 
photometric accuracy, all known sources are fitted again after examining and fitting all 
the candidate objects. The stars are then subtracted to search for possible lost objects, 
which are examined and fitted with the same procedure in the original frame. This step was 
iterated three times for each filter. 

All the sources detected in the neighborhood of the two SSCs were discarded. The 
magnitudes of all stars were then calibrated following the same procedure described 
in Section~3.2. Although StarFinder provides the stellar flux in an infinite aperture,  
we translated the fluxes into a 0$\farcs$5 aperture using aperture correction factors
in order to apply the zero points correctly. In this case the correcting factors were 
derived directly from the same PSF templates adopted for the photometric reduction. 
This accounts for the residual flux from contaminating objects present in the PSF halo 
of such a crowded stellar field. The values applied were +\,0.034 and +\,0.062~mag for 
the F110W and F160W filters, respectively.

The F110W and F160W catalogs, derived independently with the above procedure, were 
then compared and only the stars with a spatial offset smaller than 1/2 pixel were 
retained. This led to 3133 stars detected in both filters. The CMD constructed with 
this sample is shown in panel $a$ of Figure~\ref{figemi}. This CMD displays the same 
general characteristics seen in the CMD of the complete sample obtained with DAOPHOT 
(panel $a$ of Figure~\ref{cmdnic}), even without imposing any additional selection on 
the photometric error. This implies that both methods, although different, dealt quite 
successfully with the extreme crowding of this data set, and provide a qualitatively
very similar result. An additional, very restrictive selection was made on the CMD 
obtained with StarFinder in order to further corroborate this conclusion: the 
correlation coefficient threshold was set to the much higher value of 0.95. Such a 
value limits the selection to objects with an extremely high probability of being 
resolved single stars with a good S/N. The resulting {\it hyper-selected} CMD includes 
only 730 stars (panel $b$ of Figure~\ref{figemi}), but still preserves the general 
structure of the less restricted CMD.

A direct comparison of the two photometric solutions is shown in Figure~\ref{comp}, 
where the top panel refers to the F110W band and the bottom panel to the F160W band. 
A zero-point offset of $\sim$~0.05~mag affects both F110W and F160W magnitudes, in the 
sense of DAOPHOT providing fainter values than StarFinder. We investigated this offset 
at some length and reached the conclusion that it is probably mainly due to the different 
treatment of the background. Some differences have also been found in the centering and 
interpolating techniques. Given the overall uncertainties affecting the photometry, we 
consider this small offset as a confirmation of the reliability of the derived magnitudes, 
rather than an inconsistency.

Summarizing, the direct comparison between the results of StarFinder and DAOPHOT 
(Figure~\ref{comp}) suggests that the CMDs obtained with the two different packages 
(panel $a$ of Figures~\ref{cmdnic} and~\ref{figemi} for DAOPHOT and StarFinder, 
respectively) are qualitatively similar. Small systematic errors in the magnitude 
calibration of NGC~1569 photometry (the zero-point offsets) could still be present. 
The formal DAOPHOT errors underestimate the real uncertainties in  severe crowding 
conditions. This result is confirmed by the artificial star analysis (see the following 
Section~3.4), where a quantitative estimate of the real photometric errors as a function 
of magnitude will be given.

\subsection{Artificial star analysis}

The interpretation of the photometric results for crowded fields is critically 
dependent on the results of the completeness analysis. We used the DAOPHOT routine
ADDSTAR for 20 tests in both filters. For each test, no more than $\sim$5$\%$ of the 
number of stars detected in each half-magnitude bin was added as artificial stars at 
random positions in the {\it drizzled} images. This low fraction was chosen in order 
not to significantly increase the already extreme crowding. The images were then
analyzed using the same procedures applied to the science frames. The tests in the 
F110W filter were performed by creating a catalog with the automatic star detection 
routine. In F160W the positions of real + artificial stars were assumed to be known,
and used as input for the following photometric reduction. The constraints on the 
photometric errors and shape parameters were $\sigma_{\rm{DAO}}\,<\,0.1$~mag and  
$\chi^2\,<\,1.5$, $-\,0.4\,<$~{\it sharpness}~$<\,+\,0.4$ in both filters, respectively.

The effective completeness was assessed by calculating the number of stars recovered 
in each half-magnitude bin. Stars were considered to be successfully recovered if the 
magnitude difference $\Delta m $ between the input and recovered values was less than 
0.25~mag (i.e., the half width of the magnitude bin). This is the mean value for which 
an artificial star can still be considered inside the same magnitude bin. Our selection 
included stars for which the magnitude brightening was due to a negligible blend with 
one or more objects having a total flux $\lsim$1/4 of their initial value. Table~1 shows 
the completeness factors (percentage of recovered artificial stars) as a function of 
magnitude for both the F110W and F160W filters. The corresponding uncertainties were 
calculated with the method proposed by Bolte (1989). Completeness is almost 100\% for 
$m_{\rm{F110W}}$, $m_{\rm{F160W}} \lsim 17$, while it drops to $\sim$50\% around 
$m_{\rm{F110W}}=21.5$ and $m_{\rm{F160W}}=20$, and is almost 0\% for 
$m_{\rm{F110W}}\gsim23.5$ and $m_{\rm{F160W}}\gsim22$. We can assume that the photometry 
is robust down to $m_{\rm{F110W}}=21.5$, $m_{\rm{F160W}}=20$, where the completeness factor 
is better than or equal to 50\%.

In addition, we investigated how stellar blending affects the photometry. Figure~\ref{blend}
shows the magnitude difference $\Delta m$ of the recovered artificial stars as a function of
their input magnitude. It can be seen that crowding affects the photometry in two ways: some
stars are found artificially brighter because of the addition of their flux to that of a 
blending object; some are found artificially fainter, because the background is enhanced in 
this crowded field. The first effect is seen at virtually all magnitudes; the second is stronger 
for the fainter stars. In the light of this, we are inclined to interpret the differences 
between the DAOPHOT and StarFinder photometry as due to a different way to evaluate the 
background, DAOPHOT assigning a sistematically brighter background. The previous considerations 
suggest that some blends can be hidden by a bad estimate of the background, especially at fainter 
magnitudes. 

Finally we point out that the distribution of the $\sigma_{DAO}$ of the artificial stars 
is very similar to that of the observed stars. Comparing Figure~\ref{error} with 
Figure~\ref{blend}, it is clear that the true photometric error $\Delta m$ tends to be larger 
than the parameter $\sigma_{DAO}$ for a substantial fraction of objects. This happens especially 
at faint magnitudes. Most of the artificial stars appear anyway to have a small true error. This 
implies that the general interpretation of our CMD as presented in Section~4 is robust. However, 
a special care will be devoted to the modelling of the photometric errors, and in particular of 
the blending effect, in order to derive quantitative results on the SF history in this field of 
NGC~1569 via theoretical simulations. This will be done in a forthcoming paper (Tosi et al. 
2000$b$).

\section{Color-magnitude diagrams and luminosity function in the NIR}

The data presented above allow us to examine for the first time the red population 
of NGC~1569 down to quite low stellar masses, and, hence, to quite old ages. 

The CMD in the F110W and F160W bands generated with DAOPHOT is in Figures~\ref{cmdnic} 
$a$ and $b$, for the catalogs with the shape screening and  with the cut in the photometric 
error, respectively. De Marchi et al. (1997) discuss the three SSCs in this region, two 
of which form a close pair. None of the clusters is resolvable into single stars with 
these data, hence they do not appear in the CMD. A separate paper (Origlia et al. 2000) 
analyzes their integrated optical/NIR properties.

The two panels of Figure~\ref{cmdnic} demonstrate that the major features of the galaxy 
stellar population are well delineated. Obviously the diagram with the tighter selection 
criteria contains fewer stars due to the brighter magnitude limit (\mj~$\approx 23.5$ vs.
\mj~$\approx 25$) and the rejection  of extremely red objects. The CMD obtained with 
StarFinder in Figure~\ref{figemi}$a$ has almost the same number of objects as that in 
Figure~\ref{cmdnic}$a$, and the same overall features. The StarFinder CMD with the tightest 
criterion for star selection in Figure~\ref{figemi}$b$ provides the most conservative sample 
of resolved single stars. Despite its much lower number of objects (730), the latter CMD 
preserves the major features and delineates the characteristics of the red, faint population 
even better. In the following we will use the CMD in Figure~\ref{cmdnic}$b$ as the reference 
diagram for the interpretation of these data in terms of stellar populations.

A blue plume dominates the stellar distribution in the optical CMD of NGC~1569 (G98), and of 
dwarf irregulars and blue compact galaxies in general. The blue plume is usually populated 
by stars on the main-sequence (MS) and at the blue edge of the core-helium-burning phases. 
The NIR CMD is a mirror image of the optical one, with a red plume clearly dominating 
Figures~\ref{cmdnic} and \ref{figemi}. The red plume is centered at \mj~--~\mh~$\simeq1.2$ 
and extends from \mj~$\simeq18$ all the way down to the faintest magnitudes.

NGC~1569 is quite close to the Galactic plane ($b=+11^{\circ}$), and contamination by Galactic 
foreground stars could heavily affect the observed CMD. VB addressed this issue in the optical 
spectral region by analyzing ground-based observations of nearby fields, and derived a very low
contamination probability. We ran for our NIR field the Galaxy model described by Casertano, 
Ratnatunga, \& Bahcall (1990) and kindly made available by S. Casertano. It is based on a 
three-component geometry, the thin (young) and thick (old) exponential disks, and the 
de~Vaucouleurs spheroid. Extinction is included in the model calculations by taking into 
account the obscuration law of Sandage (1972). The prediction is $9\times10^4$ stars brighter 
than $V =27$ per square degree at the position of NGC~1569. $V=27$ is the faintest magnitude 
corresponding to our limiting NIR value $J\approx24$ in the most extreme case of red stars 
on the TRGB with $V-J\approx3$. This translates into about 3 Galactic stars within the field 
of view of our NICMOS observations, suggesting that contamination by foreground stars is 
negligible in the NIR as well.

The low Galactic latitude of NGC~1569 also implies a significant foreground extinction 
($E(B-V)=0.56$, Israel 1988). An additional intrinsic reddening of $E(B-V)=0.11$ is 
derived from the Balmer decrement of the ionized gas in the vicinity of the two SSCs 
(Gonz\'alez-Delgado et al. 1997). The extinction derived from the UV continuum slope is 
often smaller than that from the Balmer decrement in star-forming galaxies (Fanelli, 
O'Connell, \& Thuan 1988; Calzetti, Kinney, \& Storchi-Bergmann 1994). Origlia et al. 
(2000) give an independent estimate of the foreground extinction of $E(B-V)=0.55$ from 
the strength of the broad Galactic absorption feature at 2200~\AA. An additional intrinsic 
component of $E(B-V)=0.15$ is then derived from a comparison of the observed and theoretical  
UV continuum slopes. This confirms the total reddening of $E(B-V)=0.7$ derived by 
Gonz\'alez-Delgado et al. (1997). Devost, Roy, \& Drissen (1997) infer the same line-of-sight 
total extinction across the main body of $E(B-V)=0.7$ from the Balmer H$\alpha$/H$\beta$ line 
ratio, although only $E(B-V)=0.50$ is attributed to the Galactic component. In view of these 
uncertainties, we will assume that stars in our CMD are affected only by Galactic extinction
with $E(B-V)=0.56$. An additional intrinsic reddening of up to $E(B-V)=0.15$ would not 
change the major conclusions. This extinction would translate into  $A_{\rm J}=0.15$ and 
$E(J-H)=0.08$. Such an intrinsic component, which might be variable across the disk of 
NGC~1569, would hardly be noticeable in the CMD. 

We have overlaid the Padua stellar evolution tracks with metallicities $Z=0.004$ (Fagotto
et al. 1994$b$) and $Z=0.0004$ (Fagotto et al. 1994$a$), on the reference CMD in 
Figure~\ref{tracks}. These metallicities correspond to \ZSUN/5 and \ZSUN/50, respectively. 
The theoretical tracks have been transformed into the observational plane by adopting 
$E(B-V)=0.56$ and $(m-M)_0=26.71$ as in G98. The tables for bolometric correction and 
temperature-color conversion in the HST VEGAMAG photometric system are from Origlia 
\& Leitherer (2000). $Z=0.004$ is close to HII region abundances. Chemically consistent 
models for the evolution of galaxies (M\"oller, Fritze-v.Alvensleben, \& Fricke 1997)
predict $Z=0.004$ in stars that produce the bulk of the bolometric luminosity in very 
late-type spiral galaxies after a Hubble time. The gas abundance is as high as $Z=0.006$ 
in their closed-box model with an almost constant SF. Only a small fraction of the metals 
produced by massive stars ($\sim$30\%  for a 10$^9$~\MSUN\ galaxy undergoing a normal 
starburst) is retained by the weak gravitational potential of the galaxy dark matter halos 
if large-scale outflows are present (e.g., D'Ercole \& Brighenti 1999; Mac Low \& Ferrara 
1999). In this case the overall metallicity of the ISM could be lower than 0.006. We can 
thus conclude that tracks at a metallicity of $Z=0.004$ is appropriate for stars brighter 
than \mj~$\approx21$ in our NIR CMD of NGC~1569. This is also supported by G98's analysis 
of the young stellar population resolved with WFPC2 in the optical. The metal-poor Padua 
tracks at $Z=0.0004$ should instead better represent the older stellar population, i.e., 
the very old stars in the RGB with an age of a Hubble time. The remaining faint stars with
ages less than $\sim$~1~Gyr not in the RGB stage are expected to have a metallicity bracketed 
by the two extremes $Z=0.0004$ and $Z=0.004$.

The masses of the stars plotted in the figure range from 0.8 to 30~\MSUN~ for both 
metallicities. We have plotted all the evolutionary phases for high- and intermediate- 
mass stars, whereas for low-mass stars (i.e. $\leq$\,1.7~\MSUN) the tracks end at the 
TRGB in order to avoid confusion. The subsequent bright early AGB (E-AGB) phase does 
not become brighter than the TRGB. Thus, with the adopted distance modulus, most of 
the stars brighter than the TRGB and redder than \mj\,--\,\mh~$\approx 1.4$ should 
be TP-AGB (thermally pulsating AGB) stars. Once dereddened, this color becomes 
(\mj\,--\,\mh)$_0\approx 1.1$. This value agrees very well with the blue edge of the 
TP-AGB stars inferred in Schulte-Ladbeck et al. (1999). 

In both panels of Figure~\ref{tracks} the main-sequence corresponds to the almost 
vertical lines at $0\leq$\,\mj\,--\,\mh\,$\leq0.2$ and the turn-off is recognizable 
as a small bluer hook on these lines. Among the later evolutionary phases, we can 
clearly distinguish the almost horizontal blue loops of core-helium burning, the 
bright red sequences of the AGB of intermediate-mass stars, and the RGB of low-mass 
stars. Due to its constant luminosity, the theoretical TRGB for $Z=0.004$ stands 
out at \mj~=~22.5 and \mj\,--\,\mh~from 1.45 to 1.80, where redder colors correspond
to lower masses. This feature is fainter (\mj~=~22.75), bluer, much less extended in 
color (\mj\,--\,\mh~from 1.10 to 1.20), and is no longer vertical at $Z=0.0004$. 

The tracks with $Z=0.004$ are a significantly better representation of the overall 
features of  the resolved stellar population in the NIR. In contrast, the $Z=0.0004$ 
tracks are too blue. This applies to the faintest old stars as well, suggesting that 
even these stars have metallicity higher than $Z=0.0004$. This does not exclude the 
presence of very metal-poor old stars, but they would be hidden by the population 
of the more metal-rich, younger stars around \mj~=~22.75 and \mj\,--\,\mh~=~1.15. 
Only simulations of synthetic CMDs would allow us to constrain the metallicity of 
the stars with different ages and masses in a more quantitative way. In the following 
we will thus assume $Z=0.004$ as a reasonable approximation for the average metallicity.
 
Very few stars populate the MS in our reference CMD. The blue loops are fairly well 
populated by stars down to 7~\MSUN, but the vast majority of the stars are located in 
the very reddest part of the CMD, and the red plume turns out to be populated by red 
giants and supergiants. The AGB is well populated by intermediate-mass stars with $1.8 
\leq M/M_{\odot}\leq 9$ (see also VB), and ages ranging from $\sim\,$30~Myr to $\sim\,$1~Gyr.
The RGB is populated by low mass stars, with ages up to a Hubble time. Therefore we 
conclude that the stars in the field of this galaxy cover the whole age range from a 
few Myr to a Hubble time. There is no doubt that the RGB is sampled by our data. The 
most tightly selected stars of Figure~\ref{figemi}$b$ show  a faint tail on the cool 
side of the red plume, extending from \mj\,--\,\mh\,$\simeq$\,1.2 to 2.0. This tail
persists even when most of the faintest objects are removed by the most conservative 
selection criteria. The red colors of these stars strongly suggest they belong to the 
RGB of low-mass stars.

The theoretical TRGB is 2.5~mag brighter than our absolute faint magnitude limit 
(\mj$\approx25$) and only 1~mag brighter than the limit in the reference CMD 
(\mj$\approx23.5$). At \mj~=22.5, the formal photometric error reaches up to 
0.1~mag (see Figure~\ref{error}), but the actual uncertainty may be quite larger 
since the completeness reaches only $\sim$\,15\% and the blending probability is 
no longer negligible (see Figure~\ref{blend}). All these caveats suggest it is not 
prudent to derive the distance of NGC~1569 directly from the TRGB in our data.

In Figure~\ref{lf} we show the differential (dashed line) and the integrated 
(solid line) LF in the F110W filter for the 2423 stars in the reference CMD. 
The slope of the differential LF has been derived with a maximum likelihood 
fit. Down to \mj~=~19, where our data are essentially complete, the slope is 
$\Delta\log N/\Delta$\mj~$=1.0\pm0.1$. If we extend the fit to \mj~=~20, where 
completeness is 85\%, the slope becomes $0.63\pm0.04$. Unfortunately, the 
incompleteness at \mj~=~22.5 is too high and prevents any appearance in the 
LFs of Figure~\ref{lf} of the change of slope often used to identify the 
location of the TRGB in less distant systems.

Summarizing, the stellar distribution of our reference NIR CMD in Figure~\ref{cmdnic}$a$ 
is almost unaffected by contamination from Galactic foreground stars or by differential 
intrinsic reddening. The only relevant effects are the high photometric uncertainties and 
low completeness factors. They start to become severe at relatively bright magnitudes 
(\mj$\,\approx\,21.5$). Photometric errors and blending are probably the major causes of the 
wide spread of the stellar distribution towards fainter magnitudes. In addition, the low 
completeness factors of $\lsim$\,50\,\% play a key role in complicating the interpretation 
of our data for look-back times of more than 1~Gyr. A metallicity of $Z=0.004$ reproduces 
the overall features of the observed CMD rather well. This value agrees with measurements
in  HII regions.  The strong and bright ``red'' plume of AGB stars seen in NGC~1569 is  
directly related to the peculiar and intense SF of this dwarf irregular in the last $\sim$1~Gyr. 
The detection of very faint red stars in their RGB stage with ages up to 12~--~15~Gyr indicates 
SF activity over a whole Hubble time.

\section{Cross-correlation with the optical data}

We cross-correlated the optical and NIR catalogs obtained with WFPC2 (G98) and 
NICMOS (this work), and found a very small fraction of stars common to the four 
filters. 10 stars which were clearly visible in all the four images were selected, 
and their coordinates were used to derive the geometric bilinear coordinate 
transformation between the two catalogs. Different software tools were applied, 
with no effect on the transformation solution. We interpret the absence of a clear 
correlation between the two catalogs as due to severe blending between coinciding 
blue and red stars. 

Support for this explanation  is provided in Figure~\ref{spectra}, where we 
superimposed the spectral energy distribution of stars with different spectral 
types on the transmission profiles of the optical and NIR filters. Each stellar
spectrum is normalized to a magnitude of 22 in the HST VEGAMAG system for the 
F555W filter. If a blue B0 and a red M0 star spatially overlap, the only one visible 
in the NIR bands will be the later spectral type. The bluer counterpart will be 3 (4) 
mag fainter, i.e., the flux will be 1/20 (1/40) lower, in F110W (F160W). These two 
stars have a difference of $\sim$2~mag in the bluest F439W filter, where now the bluer 
star has a flux 6 times higher than the red star. However, the blend of two stars with 
the same flux in F555W can be easily recognized due to cross-correlation between the 
two optical filters. Therefore this extreme case should not apply to our CMD. Now we 
suppose that these two stars have slightly different magnitudes in F555W, for example 
the B0 star being 1.5~mag brighter than the M0 star (flux ratio $\sim$~1/4). The flux 
of the B0 star dominates in both optical filters, with the magnitude difference in F439W 
increasing to $\sim$3.5~mag. In this second example the difference between the B0 and M0 
star magnitudes decreases to 1.5 (2.50) mag in F110W (F160W), but remains still quite 
high. This implies that only the M0 star will be visible in the two NIR bands, and only 
the B0 star will be detected in the optical filters. This explains the absence of a 
correlation between the optical and NIR catalogs.

The blending is particularly severe for NGC~1569 due to the high crowding, as well as 
by the use of an optical filter at relatively short wavelength (F439W). As a result, 
the NIR is biased against the detection of very blue stars. In addition, the peculiar 
SF history of this dwarf irregular galaxy plays a key role. It has recent strong SF 
activity  coupled with a rather intense SFR at intermediate look-back times. 
Schulte-Ladbeck et al. (1999) showed that for less crowded fields, like that of 
VII~Zw~403, and with the use of redder optical filters (F555W and F814W), the 
cross-correlation between WFPC2 and NICMOS data is much more straightforward than 
in NGC~1569. The excellent cross-correlation of the stars resolved in the F555W, F814W, 
F110W, and F160W images of the starburst galaxy NGC~1705 supports this interpretation
(Tosi et al. 2000$a$).

The lack of overlap between the two catalogs indicates that the optical and NIR CMDs
are anti-correlated: the optical CMD shows the very blue (young) stellar population 
and the NIR CMD the reddest (oldest) objects. These two CMDs are complementary, once 
the completeness factors are taken into account. In Figure~\ref{compl_color} we show 
the completeness factor in F160W as a function of the stellar color for different 
fixed values of the F110W magnitude. This figure serves to illustrate that completeness 
is not only a function of the magnitude but also of the color. For example, at 
$m_{\rm{F110W}}\approx21$, just 0.5~mag above the turn-off of a 30~\MSUN~star, objects 
in the blue plume with $m_{\rm{F110W}}-m_{\rm{F160W}}\approx 0.1$ have a completeness 
factor of $\sim$~20\%. This value increases to $\sim$~55\% for stars in the red plume 
with $m_{\rm{F110W}}-m_{\rm{F160W}}\approx1.2$. In other words different portions of 
the CMD have different completeness factors, indicating that the blue plume has a worse 
sampling in the NIR by up to more than a factor of 2 than the red one. We conclude that 
each CMD by itself is suitable to infer the SF history over the whole look-back time, 
even if different types of stars are sampled.

\section{NIR stellar content of NGC~1569 and its starburst properties}

We investigated the spatial distribution of the resolved stars as a function of their 
age in order to correlate the starburst properties of NGC~1569 with its stellar population.

We grouped stars in the NIR CMD into three different mass ranges and related them to an age 
interval by means of the reference Padua tracks at $Z=0.004$. Young stars are defined as 
objects with initial mass greater than 8~\MSUN, i.e., an age less than $\sim\,$50~Myr. 
Intermediate-age stars have an initial mass in the range $1.9\lsim M \lsim8$~\MSUN, 
corresponding to an age interval from $\sim\,$50~Myr to $\sim\,$1~Gyr. Old stars are all 
the objects with a mass smaller than 1.9~\MSUN, and an age greater than $\sim\,$1~Gyr. 
Figure~\ref{tracks_distr} illustrates the relevant portions of the NIR CMD. In order 
to avoid confusion between tracks and resolved stars, we omitted the observational points 
(see Figure~\ref{tracks}). We distinguished between young and intermediate-age stars by 
approximating the lower envelope of the 9~\MSUN~ stellar evolution track with two straight 
lines, which are connected at the reddest part of the blue-loop stage. The boundary between 
intermediate-age and old stars is indicated by a straight line along the 1.9 ~\MSUN~stellar 
evolution track. One should keep in mind that metallicity decreases with decreasing stellar 
initial mass, and that more metal-poor tracks in the observational plane become bluer. We 
already discussed the metallicity issue in Section~4. Here we want to recall that the 
appropriate metallicity for the reddest stars in the RGB phase should not be very different 
from the adopted $Z=0.004$ value, but that the oldest more metal-poor RGB stars are probably 
in the region around \mj~$\approx$~22.75 at bluer colors (\mj~--~\mh\,$\approx$\,1.1), and 
contaminate the intermediate-age stellar population of this area. 

In Figure~\ref{distr} we show how the different classes of stars are distributed in our 
NIC2 images. The main result is a degree of segregation: the youngest stars are more 
clustered around the two SSCs (top left panel), the intermediate-age objects have an almost 
uniform distribution (top right panel), and the oldest stars are located in the outskirts 
of the starbursting region (bottom left panel). The absence of old stars in the innermost 
regions is probably caused by crowding, as all stars fainter than \mj~=~22.5 (both blue 
and red) are resolved only in the regions with low star density. The bottom right panel of 
Figure~\ref{distr} contains all the resolved stars, independently of their age. We conclude 
that the very young high-mass stars are effectively concentrated in the core of the 
starbursting region, while stars older than 50~Myr constitute a more uniform population 
underlying the most recent starburst. 

This finding strengthens the idea of a recent starburst ($<$\,50~Myr) in NGC~1569 
located around the two central SSCs, which triggered the large galactic outflow 
seen at other wavelengths (see Section~1). The SSCs are the results of a violent and 
concentrated SF episode in the last $\sim$~10~Myr (Gonz\'alez-Delgado et al. 1997; 
Origlia et al. 2000). The large number of young massive stars in the field around these 
SSCs resolved by us and by G98, as well as the  Wolf-Rayet population detected from their 
HeII emission by Buckalew et al. (2000), possibly indicate a different and independent 
SF mode more extended both in time and space. Alternatively, the field stars could have 
been ejected by the two SSCs, or left behind after clusters with ages more than 5~--~10~Myr 
have dissolved. We estimated the mass of gas converted into stars over the last few tens 
of Myr in the central field of view of NGC~1569. We included 549 stars in the region defined 
as {\it young} in Figure~\ref{tracks_distr}. These are the descendents of MS stars with mass 
greater than $\simeq$\,9\,\MSUN. Scaling this number by the ratio between the post-MS to total 
stellar lifetimes ($\sim$\,0.06), the sampled stars correspond to $\sim$9000 objects with mass 
greater than 9\,\MSUN, i.e., age younger than 40~Myr. Assuming a Salpeter IMF ($\alpha=2.35$), 
from 0.1 to 120~\MSUN, the corresponding converted mass is $\sim 1.4 \times 10^{6}$~\MSUN, much 
higher than the $3.3\times 10^{5}$~\MSUN~ of SSC A (Ho \& Filippenko 1996). Transforming this 
mass in the SFR per unit area in our NIC2 field of view, we derive 
$\sim$\,1\,\MSUN~yr$^{-1}$\,kpc$^{-2}$. Taking into account the incompleteness of our data shown 
in Figure~\ref{compl_color}, this rough estimate compares well with the equivalent SFR of 
4~\MSUN~yr$^{-1}$\,kpc$^{-2}$ derived by G98.

\section{NICMOS photometry of NGC~1569: a challenging case}

HST NICMOS observations of the field of view centered around the two SSCs of NGC~1569 
constituted a challenging case from the point of view of the photometric reduction. The 
major factors affecting our analysis are the severe crowding, and the faint limiting 
magnitude sampled by our data. These two aspects are both important for NGC~1569 due to 
a combination of different physical properties of this galaxy, all acting in the same bad
direction of complicating the reduction procedure. 

The first of these parameters is an unfavorable balance between the distance and the IMF. 
This dwarf irregular galaxy is the closest example of a starbursting system outside the 
Local Group ($D=2.2$~Mpc). From an observational point of view, considering a fixed 
limiting apparent magnitude, the detection of stars with a certain absolute magnitude 
should be easier. A shorter distance increases also the range of absolute magnitudes 
sampled in the CMDs, implying the detection of fainter objects with respect to the 
brightest resolved stars. Fainter objects are less massive, thus more numerous once the 
IMF has been taken into account. This works in favor of a higher degree of crowding, with 
more difficulties in the spatial resolution of single stars. I~Zw~18 for example, located 
at a much larger distance ($D=12.6$~Mpc), does not present the same crowded field of view 
when resolved with HST NICMOS (\"Ostlin 2000). In this case only the few brightest stars 
are resolved out of the much higher number of unresolved objects.

Another factor heavily affecting the NIR observations of NGC~1569 resolved stellar population 
is the high foreground reddening of $E(B-V)=0.56$ due to our own Galaxy. The extinction makes
the apparent magnitude of the detected stars fainter. The reddening effects in the NIR are not 
as dramatic as in the optical bands, but in the case of NGC~1569 they make the magnitudes fainter 
by a factor around 0.55 and 0.27~mag in F110W and F160W, respectively. Moreover reddening can 
contribute to complicate the theoretical interpretation of the NIR CMDs, if it has a differential 
intrinsic component. In Section~4 we have, however, demonstrated that this component should be 
negligible in the NIR for NGC~1569. 

In addition, the contribution of the ionized gas can act to modify the real magnitude of the
detected stars, especially in a galaxy like NGC~1569 with a high gas content. The presence of 
ionized gas is clearly indicated by the strong, extended and structured H$\alpha$ emission 
detected in the WFPC2 F656N filter by Hunter et al. (2000). The emission line contamination 
in the F110W band is due to the presence of both Paschen-$\gamma$ and Paschen-$\beta$, while 
in F160W the only line affecting observations is Paschen-$\alpha$. This contamination is correctly 
subtracted with a good estimate of the local background during the PSF-fitting photometry 
procedure. We have seen in Section~3 how this step is affected by some problems in the case of 
NGC~1569 NICMOS data. We thus cannot evaluate which is the contribution of this physical property 
to our challenging case. 

Last but not least, the most intriguing factor adding difficulty to the photometric reduction 
of NGC~1569 is the SF history of this galaxy, an unfavorable balance between the age of different 
stellar populations and the SFR at the corresponding epochs. In general the brightest, youngest 
objects are more easily detectable, and render more difficult the detection of the faintest oldest 
ones underneath them. Examples of young single age stellar populations observed with HST NICMOS 
are the two young massive stellar clusters (Quintuplet cluster and Arches cluster) near the Galactic 
center. Figures~1 and 2 of Figer et al. (1999) demonstrate how the field of view of these two targets 
is much less crowded than our images, despite the unfavorable line of sight, due to the absence  of 
stars older than 2~--~4~Myr in the clusters. The number of stars of a certain age depends instead 
on the SFR of the galaxy at the corresponding look-back time. NGC~1569 is a special case from this 
point of view. Its SFR per unit area is the highest ever observed in the local universe, all the 
other resolved dwarf galaxies having a much lower value for this parameter (see also Section~1). 
Not only the current SFR is very high in NGC~1569, but also the one inferred on a relatively long 
look-back time scale: the very strong SF event in the last 0.1~Gyr (G98), and the milder episode 
between 1.5 and 0.15~Gyr (VB), imply many red supergiants and AGB stars which spatially cover the 
faintest oldest objects. VII~Zw~403 and Mrk~178 are other two blue compact dwarfs that have been 
resolved by HST NICMOS (Schulte-Ladbeck et al. 1999, 2000$b$) in F110W and F160W. Their NIR images, 
from which it is possible to derive a CMD with a well populated RGB, are clearly much less crowded 
than our NGC~1569 field of view. This is not only related to their much lower (about a factor of 
15) present-day SFR with respect to NGC~1569 (Roye \& Hunter 2000), but also to their different SF 
histories.

Summarizing, the extremely high SFR in the dwarf irregular galaxy NGC~1569 substained for a long 
period of time (at least 0.1~Gyr) is the main physical factor that makes it special and causes  
severe crowding.  Other blue compact dwarf galaxies, like IC~10 and I~Zw~36, have a recent SFR 
per unit area derived by H$\alpha$ just a factor of $\sim$3 lower (Roye \& Hunter 2000), and
other irregulars have the same value of SFR (NGC~4449 and NGC~5253), or a SFR just a factor of 5 
lower (NGC~4214) than NGC~1569 (Hunter \& Gallagher 1986). In all these objects, however, we do not 
expect the same photometric reduction problems if observed with HST NICMOS, unless their SF history 
implies a very high SFR not only at present, but also in more distant epochs.

\section{Conclusions}

We have obtained HST NICMOS photometry of NGC~1569 with the goal of studying
its intermediate -- old stellar population. Due to the severe crowding 
conditions and the critical technical features of the NICMOS cameras, we have 
paid particular attention to the data reduction procedures, using all the most 
updated methods to overcome the various problems affecting the data. The 
photometric uncertainties are still high, but the fact of very similar CMDs 
resulting from two different photometric packages (cf. Figure~\ref{cmdnic} and 
Figure~\ref{figemi}), suggests that the overall statistical properties of the 
resolved stellar population have been well characterized. The differences in the 
individual stellar magnitudes of the two alternative photometries (Figure~\ref{comp}), 
as well as the artificial star tests (Figure~\ref{blend}), indicate some difficulties 
in a good estimate of the background during the PSF-fitting procedure, and actual 
photometric errors significantly larger than the formal ones. 

The severe crowding has prevented us from finding a statistically significant 
fraction of stars common to both the optical (G98) and NIR (this paper) catalogs, 
due to a severe blending between blue and red stars spatially coinciding with each 
other. Nevertheless, we have demonstrated how the optical and NIR CMDs are 
complementary to each other, and suitable to infer the SFH over the whole look-back 
time, once the completeness factors are correctly inserted into the simulations.

The NICMOS photometry of NGC~1569 has turned out to constitute an especially challenging 
case, because of the extreme crowding conditions of this stellar system. We have found the 
major reason to be the exceptionally high SFR over a long look-back time ($\tau\la$\,0.1\,Gyr). 
This implies that all starbursting galaxies with a similar SF history could pose the same
photometric challenge if observed with HST in the NIR.

The NIR CMDs presented here clearly show that NGC~1569, in addition to the outstanding 
young population examined in detail by G98, contains a conspicuous old stellar component, 
with a well populated AGB (already discovered by VB) and a sufficiently delineated RGB.
Hence, the stars in the field of this galaxy cover the whole age range from a few Myr 
to a Hubble time. The metallicity of the reddest TRGB stars is in better agreement with 
Padua tracks at $Z=0.004$ than with the set at lower metal content ($Z=0.0004$). We 
cannot exclude the presence of very metal-poor RGB stars, embedded in the stellar 
distribution of the more metal-rich objects around \mj~=~22.75 and \mj~--~\mh~=~1.15. The 
youngest ($\tau<$\,50~Myr) stars seem preferentially concentrated around the two SSCs, 
while the older field stellar population of NGC~1569 has probably a more uniform distribution. 
The SFR per unit area derived from our NIR CMD for the field stars in the most recent SF 
period (1~M$_{\odot}$\,yr$^{-1}$\,kpc$^{-2}$) is consistent with the value quoted in G98, 
once completeness factors are taken into account. The estimated total mass converted into 
stars ($\sim 1.4\times 10^6$~\MSUN) is about a factor of 5 higher than that found in SSC~A. 

The intensity of the past SF activity has presumably been quite lower than
the recent one (0.5~--~3 M$_{\odot}$yr$^{-1}$kpc$^{-2}$), since at the latter 
rate G98 estimated that the system would run out of gas in only $\sim$~0.7~Gyr.
Also Gallagher et al. (1984) have estimated a SFR that is a factor of $\sim$\,4 
lower than the present one for look-back times greater than 0.5~Gyr. It is also 
possible that NGC~1569 has not experienced an almost continuous SF, but a bursting 
one, with active periods separated by quiescent intervals. To evaluate more 
precisely at what epochs the SF has actually occurred and with what intensity, 
it will be necessary to apply to these data the synthetic CMD method described by 
Tosi et al. (1991) for ground-based data and updated by G98 for HST data. The 
results of this application will be described in a forthcoming paper (Tosi et al. 
2000$b$).

As discussed in Section~4, our data at the faintest magnitudes are too incomplete and
uncertain to put stringent constraints on the galaxy distance modulus. Nonetheless, 
it is apparent that we can exclude distance moduli much different from the adopted 
value of 26.71~mag. For instance, a modulus either 1~mag shorter or 1~mag longer 
would make the distribution of the theoretical tracks quite inconsistent with the 
observational CMD.

NICMOS data of the dwarf irregular NGC~1569 clearly reveal that this galaxy contains 
an intermediate age -- old stellar population ($\tau\,>$\,1~Gyr). This result is 
consistent with the conclusions reached by many authors on the ages of the stellar 
populations in dwarf irregular and blue compact dwarf galaxies. Deep imaging in the 
NIR (Thuan 1983) and fitting surface brightness profiles in the optical (Papaderos 
et al. 1996$a$, $b$) demonstrated that unresolved blue compact dwarf galaxies usually 
have an underlying low-surface brightness component with elliptical isophotes and 
redder colors, suggesting an older evolved stellar population formed prior to the 
present starburst. This supports the idea that they are not truly primordial galaxies, 
but older systems undergoing transient and/or low level SF episodes. Their older stellar 
component has been resolved into single stars only recently thanks to the capabilities 
of HST. A classical example is I~Zw~18, observed both in the optical with WFPC2 (Aloisi 
et al. 1999) and in the near-infrared with NICMOS (\"Ostlin 2000): in this case a stellar 
component as old as at least 3~--~5~Gyr has been found. VII~Zw~403 and Mrk~178 
(Schulte-Ladbeck, Crone, \& Hopp 1998; Schulte-Ladbeck et al. 1999, 2000$b$) are other 
dwarf galaxies where HST has been able to evidence such an intermediate age -- old 
stellar population, by revealing their red and old RGB tip stars.

The general scenario emerging from all these studies is that nearby dwarf irregulars
are not genuine young stellar systems, local counterparts of the primeval galaxies 
present in the early universe (Schulte-Ladbeck et al. 2000$a$). If we want to find 
primordial systems in our local universe we have probably to look elsewhere.

\acknowledgments

We thank M. Bellazzini, E. Bergeron, H. Bushouse, D. Calzetti, S. Casertano, 
M. Dickinson, S. Holfeltz, B. Monroe, P. Montegriffo, M. Sirianni, and R. 
van der Marel for help and suggestions to overcome the data reduction and 
photometric problems. G. De Marchi, and E. Telles are also acknowledged for 
helpful scientific discussion. Support for this work was provided by NASA 
through grant GO-07881.01-96A from the Space Telescope Science Institute, 
which is operated by the Association of Universities for Research in Astronomy, 
Inc. for NASA under contract NAS5-26555. This work was partially supported also 
by the Italian ASI, through grants ARS-96-70 and ARS-99-44. Funding for A.A. 
were partly provided through an Italian CNAA fellowship. E.D. was partly supported 
by the Italian Ministry for University and Research (MURST) under grant Cofin 
98-02-32. A.A. thanks in particular the Bologna Observatory for hospitality and 
financial support during part of this work. A very preliminary version of the 
results related to this work is in A.A. Ph.D. thesis.

\clearpage

\begin{deluxetable}{ccccc}
\tablecolumns{5} 
\tablewidth{0pt}
\tablecaption{Completeness factors (percentage of recovered artificial stars) and 
relative errors as a function of magnitude for DAOPHOT}
\tablehead{
\colhead{Magnitude} &
\colhead{} &
\colhead{F110W} &
\colhead{} &
\colhead{F160W}
}

\startdata
  
    $<$16.5    & &     100     & &    100    \\
 16.5 -- 17.0  & &     100     & &  95$\pm$5 \\
 17.0 -- 17.5  & &     100     & &  93$\pm$3 \\
 17.5 -- 18.0  & &  90$\pm$7   & &  88$\pm$3 \\
 18.0 -- 18.5  & &  95$\pm$3   & &  89$\pm$3 \\
 18.5 -- 19.0  & &  90$\pm$4   & &  80$\pm$3 \\
 19.0 -- 19.5  & &  91$\pm$2   & &  68$\pm$3 \\
 19.5 -- 20.0  & &  85$\pm$3   & &  56$\pm$3 \\
 20.0 -- 20.5  & &  79$\pm$3   & &  43$\pm$2 \\
 20.5 -- 21.0  & &  71$\pm$3   & &  20$\pm$2 \\
 21.0 -- 21.5  & &  60$\pm$3   & &  10$\pm$1 \\
 21.5 -- 22.0  & &  41$\pm$2   & &   4$\pm$1 \\
 22.0 -- 22.5  & &  23$\pm$2   & &     0     \\
 22.5 -- 23.0  & &   7$\pm$1   & &    ...    \\
 23.0 -- 23.5  & & 0.4$\pm$0.3 & &    ...    \\
 23.5 -- 24.0  & &     0       & &    ...    \\

\enddata
\end{deluxetable}

\clearpage
\begin{figure}
\vspace{10truecm}
\caption[fig]{Combined drizzled image of NGC~1569 on a logarithmic scale in 
the F160W filter. The total exposure time is $\sim$85 min. The displayed 
field of view corresponds to 20$\farcs$6\,$\times$\,20$\farcs$6, with a 
pixel size in the resampled image of 0$\farcs$0375 per pixel. Orientation
is shown in the figure, with north indicated by the arrow and east by the 
perpendicular line. Clusters NGC~1569--A and NGC~1569--B have been labelled.
\label{imageH}}
\end{figure}

\clearpage
\begin{figure}
\epsscale{0.58}
\plotone{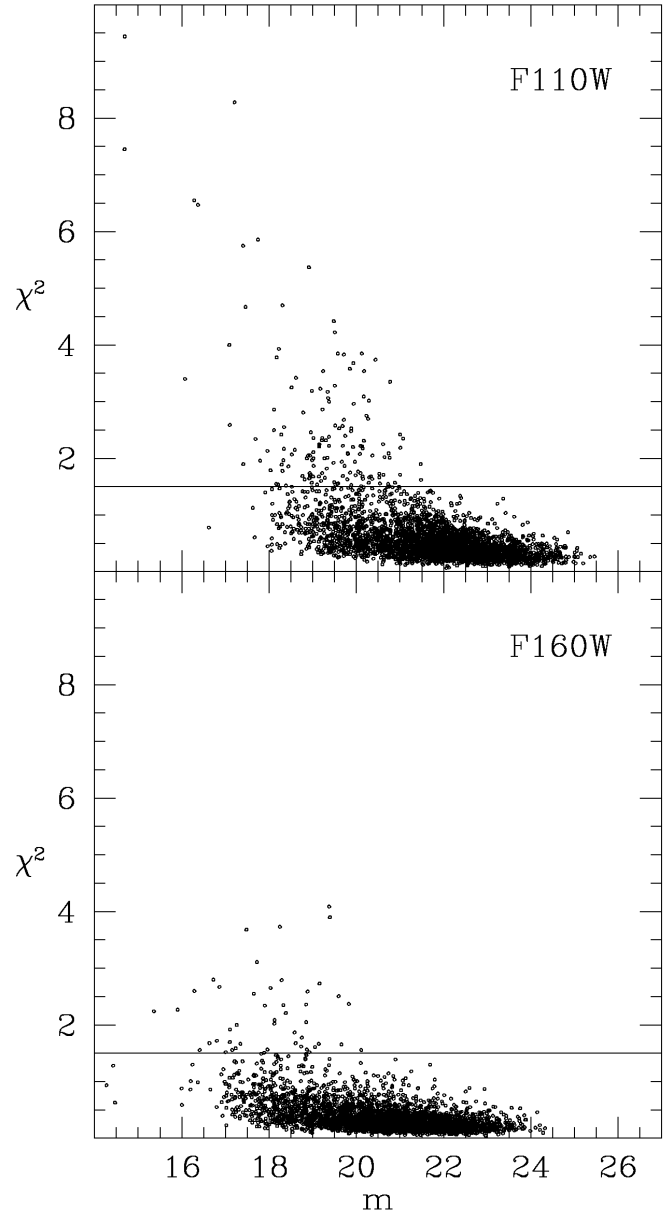}
\caption[fig]{Shape parameter $\chi^2$ as a function of the calibrated magnitude 
in both F110W (top panel) and F160W (bottom panel) filters for all the 3492 
stars retained after the photometric reduction. The horizontal line at 
$\chi^2$\,=\,1.5 in both plots represents the cut applied to our data.
\label{chi}}
\end{figure}

\clearpage
\begin{figure}
\epsscale{0.7}
\plotone{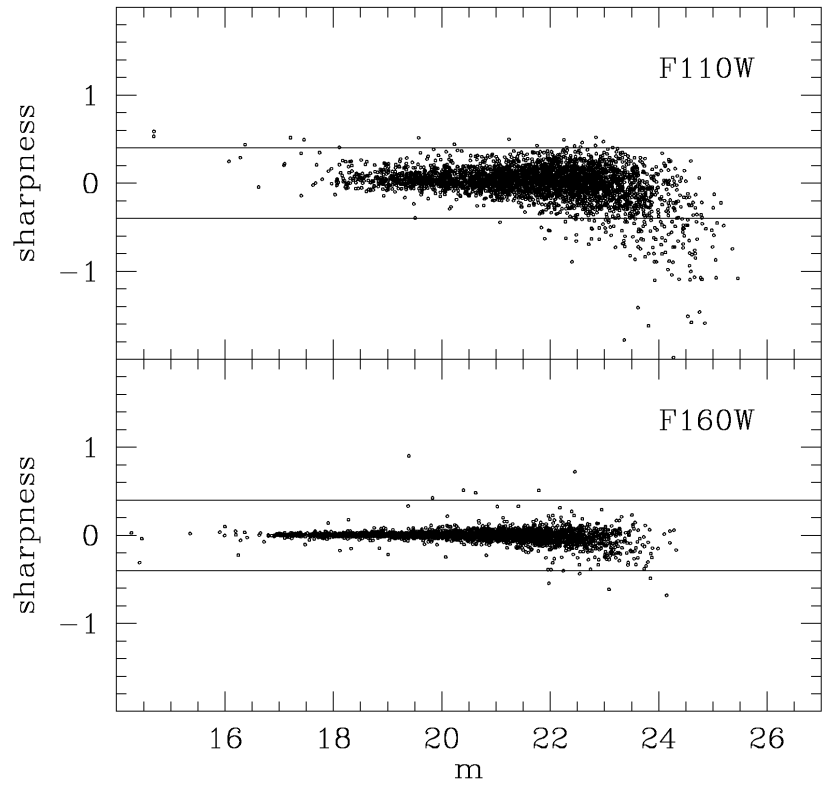}
\caption[fig]{Same as Figure~\ref{chi}, but for the shape parameter 
{\it sharpness}. Only the stars within the two horizontal lines at 
{\it sharpness}\,=\,--\,0.4 and +\,0.4 in both filters were considered 
for our final CMD.
\label{sharpness}}
\end{figure}

\clearpage
\begin{figure}
\epsscale{0.68}
\plotone{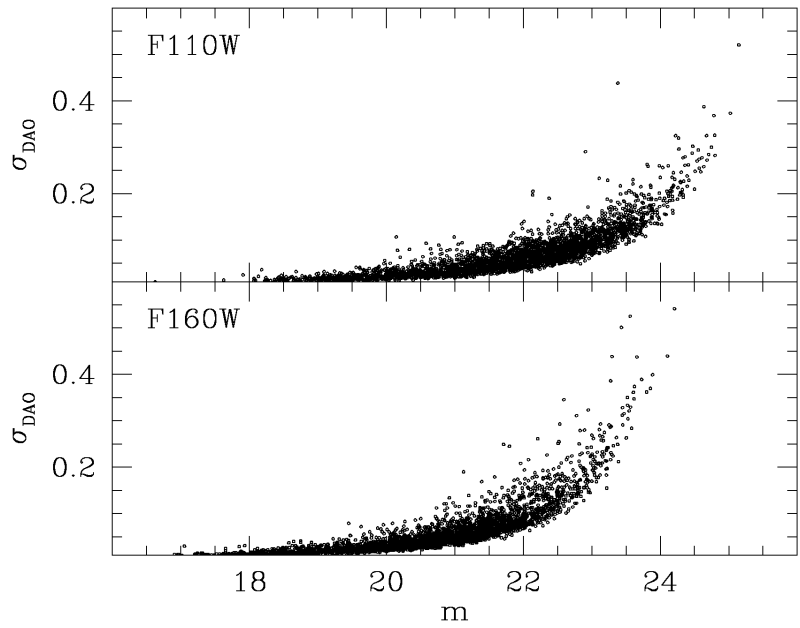}
\caption[fig]{Formal photometric errors vs. calibrated magnitude as obtained 
by DAOPHOT in both F110W (top panel) and F160W (bottom panel) filters. Only
the 3177 stars retained after the shape screening were considered.
\label{error}}
\end{figure}

\clearpage
\begin{figure}
\epsscale{0.4}
\plotone{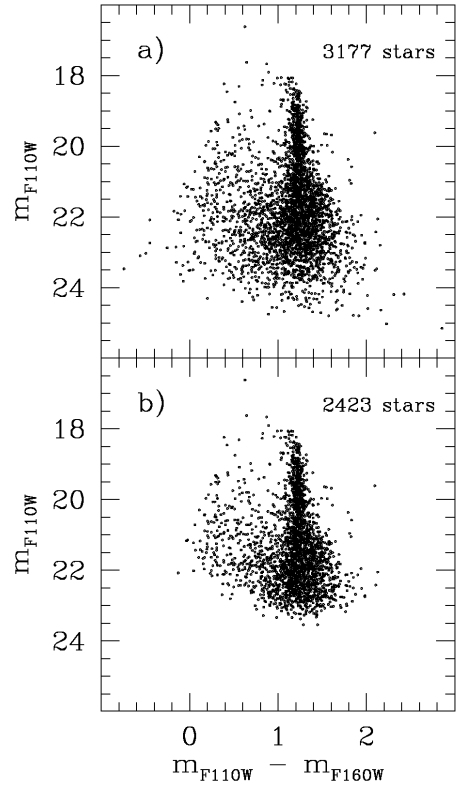}
\caption[fig]{\mj\ vs. \mj~--~\mh\ CMD of NGC~1569 in the field of the NIC2 camera. 
Photometry performed with the DAOPHOT package. Panel $a$: 3177 stars with 
$\chi^2<$\,1.5 and --\,0.4\,$<\,sharpness\,<\,$+\,0.4 in both 
F110W and F160W bands. Panel $b$: subsample of 2423 objects with 
$\sigma_{\rm{DAO}}<\,$0.1 in both filters.  
\label{cmdnic}}
\end{figure}

\clearpage
\begin{figure}
\epsscale{0.4}
\plotone{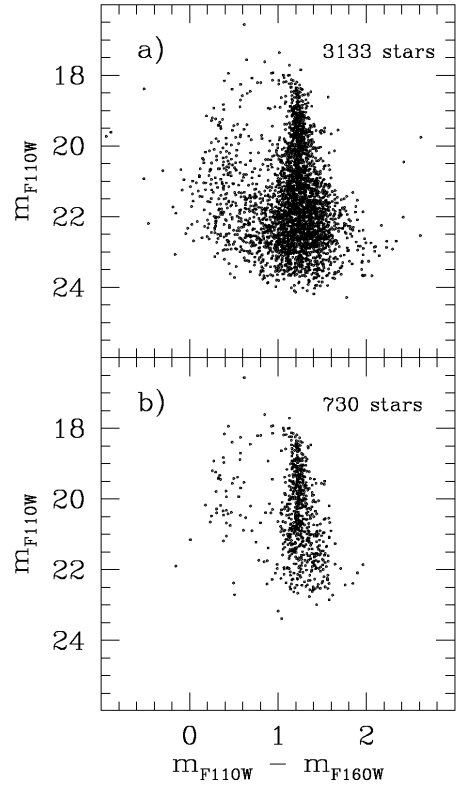}
\caption[fig]{\mj\ vs. \mj~--~\mh\ CMD of NGC~1569 in the field of the NIC2 camera. 
Photometry performed with the StarFinder package. Panel $a$: 3133 stars with 
peaks in the F110W and in the F160W images coinciding within a 1/2 pixel radius.
Panel $b$: subsample of 730 objects with PSF-fitting correlation coefficient 
higher than 0.95.
\label{figemi}} 
\end{figure}

\clearpage
\begin{figure}
\epsscale{0.5}
\plotone{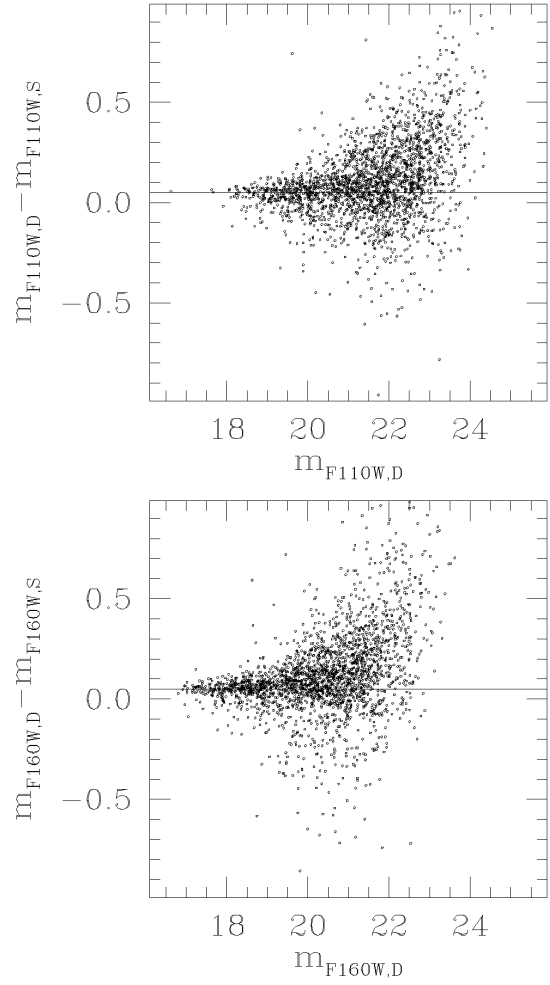}
\caption[fig]{Comparison of the DAOPHOT and StarFinder photometries in the
F110W (top) and F160W (bottom) filters for the stars in common to the
samples shown in the CMDs of panels $a$ of Figure~\ref{cmdnic} and 
Figure~\ref{figemi}. The solid horizontal lines indicate the 0.05 mag 
offset between the two photometries.
\label{comp}}
\end{figure}

\clearpage
\begin{figure}
\epsscale{0.75}
\plotone{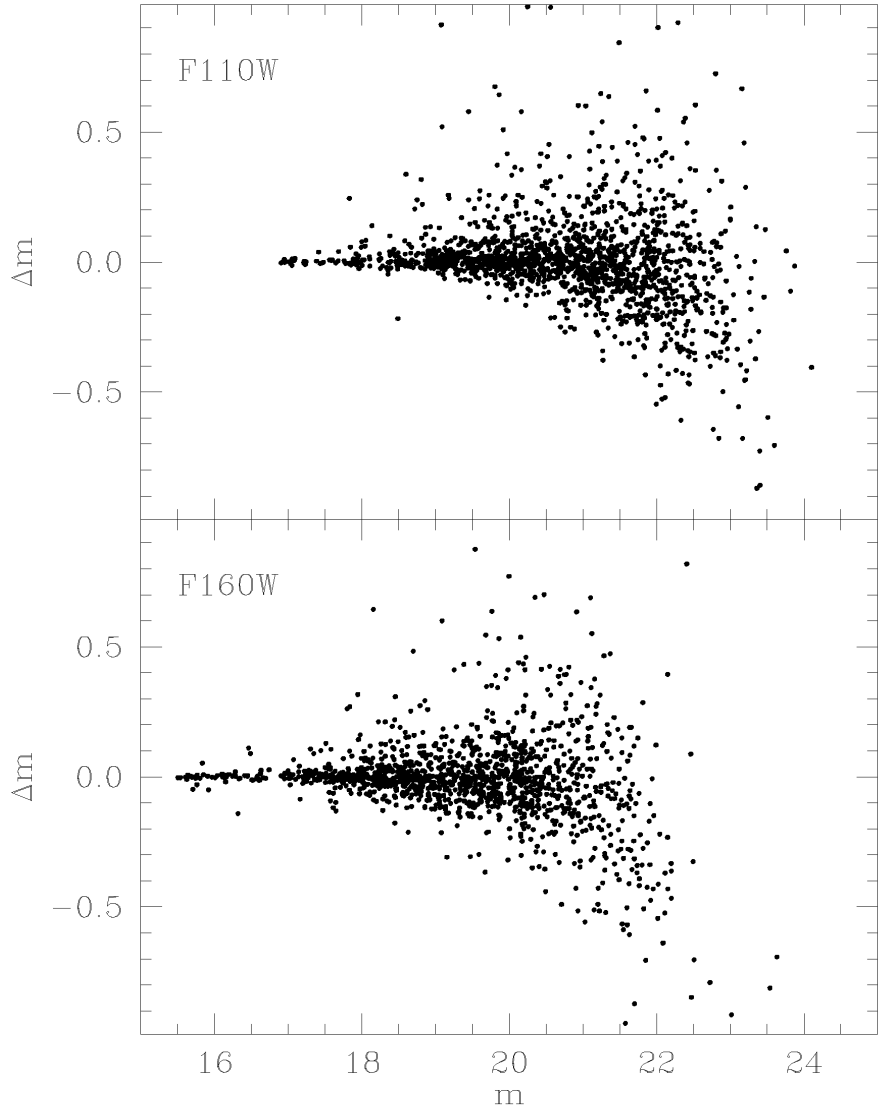}
\caption[fig]{$\Delta m = m_{\rm input} - m_{\rm output}$ vs. input magnitude obtained with 
DAOPHOT in both F110W (top panel) and F160W (bottom panel) filters for the artificial stars 
of the completeness tests. Only objects after the same shape and error screening of real 
stars are considered.
\label{blend}}
\end{figure}

\clearpage
\begin{figure}
\epsscale{0.55}
\plotone{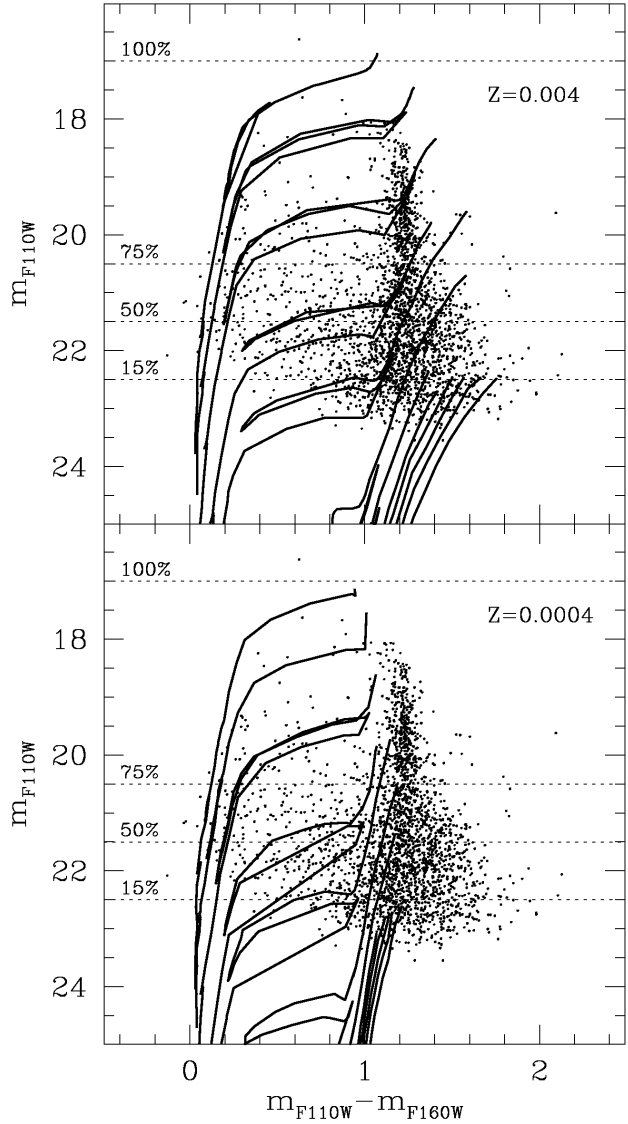}
\caption[fig]{Blow up of Figure~\ref{cmdnic}$b$, with the Padua tracks at the metallicity 
$Z=0.004$ (top panel) and $Z=0.0004$ (bottom panel) superimposed on the reference CMD. The 
stellar masses of the shown tracks are, from left to right: 30, 20, 12, 7, 5, 3, 2, 1.6, 1.4, 
1.2, 1.0, 0.8 M$_{\odot}$, corresponding to 7, 10, 21, and 56 Myr, 0.1, 0.4, 1.1, 1.8, 2.6, 
4.4, 8.5, and 20 Gyr for $Z=0.004$, and to 7, 11, 21, and 54 Myr, 0.1, 0.3, 0.9, 1.5, 2.2, 
3.6, 6.8, and 15 Gyr for $Z=0.0004$. The horizontal dotted lines give an indication of the 
completeness factors as a function of F110W magnitude.
\label{tracks}}
\end{figure}

\clearpage
\begin{figure}
\epsscale{0.8}
\plotone{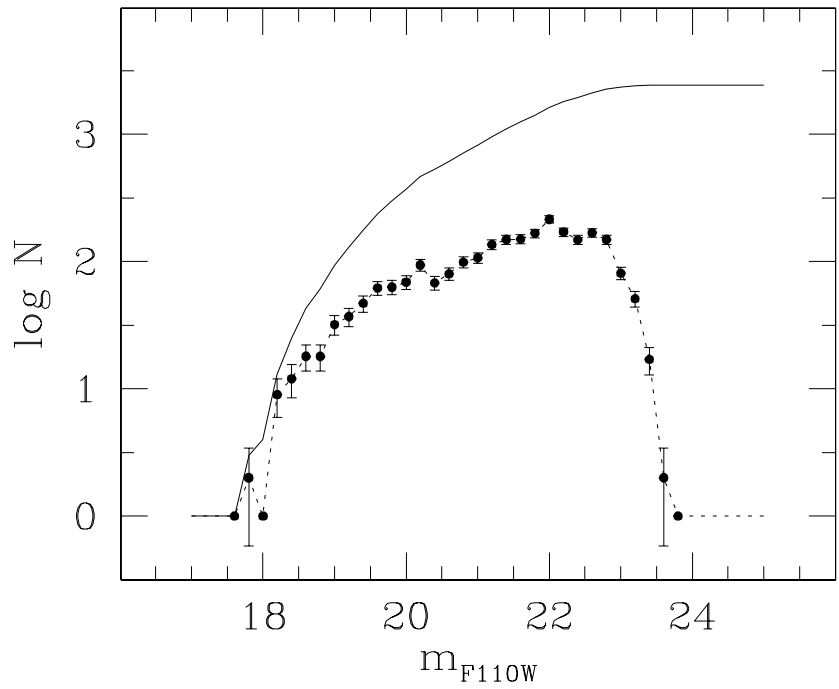}
\caption[fig]{Luminosity function in the F110W filter for the 2423 stars of NGC~1569 of 
the reference CMD. Magnitude bins of 0.2 have been considered. The dashed line, with 
points superimposed to indicate the error bars, shows the differential LF and the solid 
line the integrated LF.
\label{lf}}
\end{figure}

\clearpage
\begin{figure}
\epsscale{0.9}
\plotone{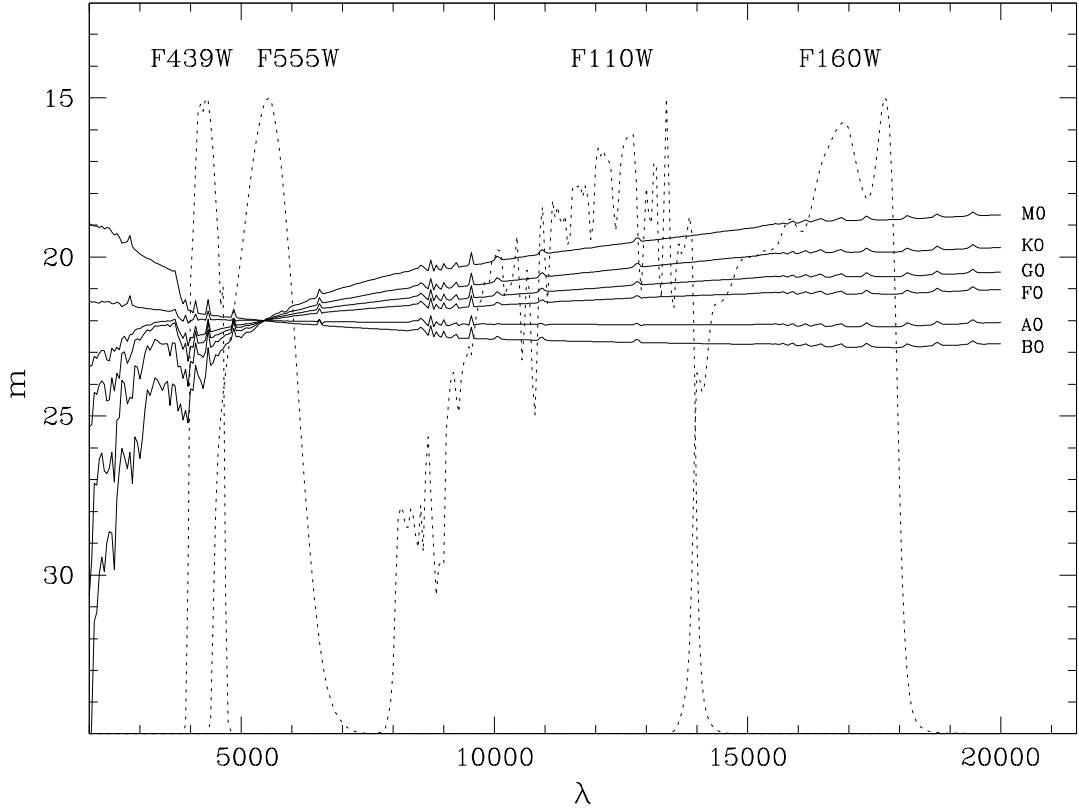}
\caption[fig]{Throughput of the WFPC2 and NICMOS filters superimposed to the 
spectral energy distribution of MS stars with different spectral type. Filters 
are normalized to the maximum value of their throughput. Stellar spectral types 
are indicated on the right, and their order is reversed on the left side of the 
figure (from top to bottom B0, A0, F0, G0, M0, and K0, respectively). No reddening 
has been taken into account, acting in the same way for each spectrum in each 
filter (the amount is the same for each spectral type).
\label{spectra}}
\end{figure}

\clearpage
\begin{figure}
\epsscale{0.7}
\plotone{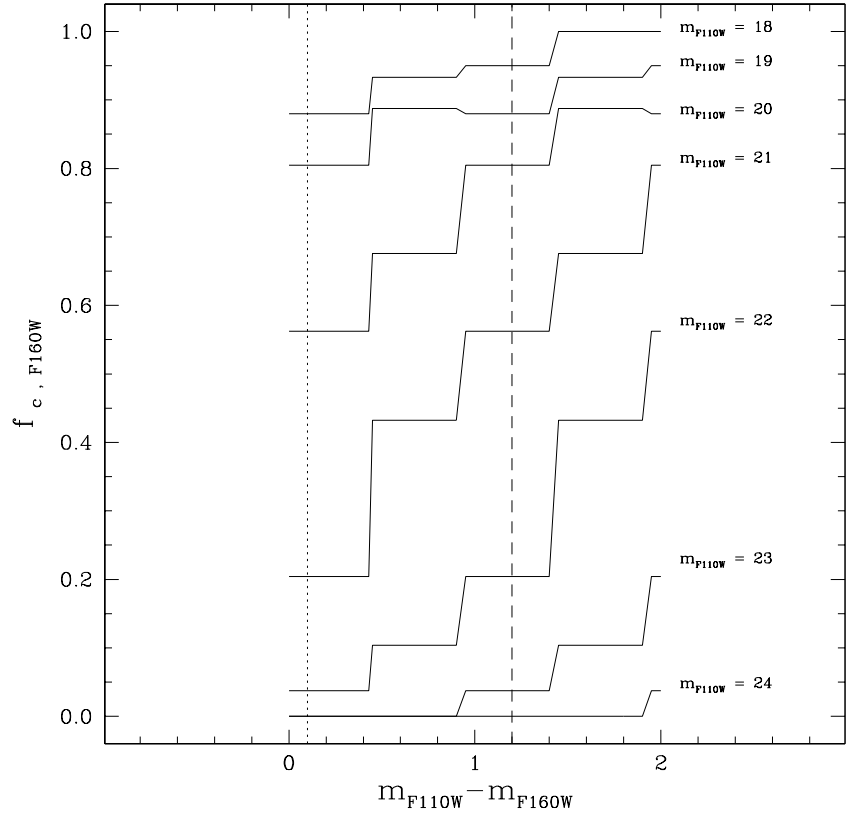}
\caption[fig]{Completeness factors in F160W as a function of the stellar color 
for fixed values of the magnitude in F110W. The vertical dotted and dashed 
lines are the color locations of the MS and red plume, respectively.
\label{compl_color}}
\end{figure}

\clearpage
\begin{figure}
\epsscale{0.7}
\plotone{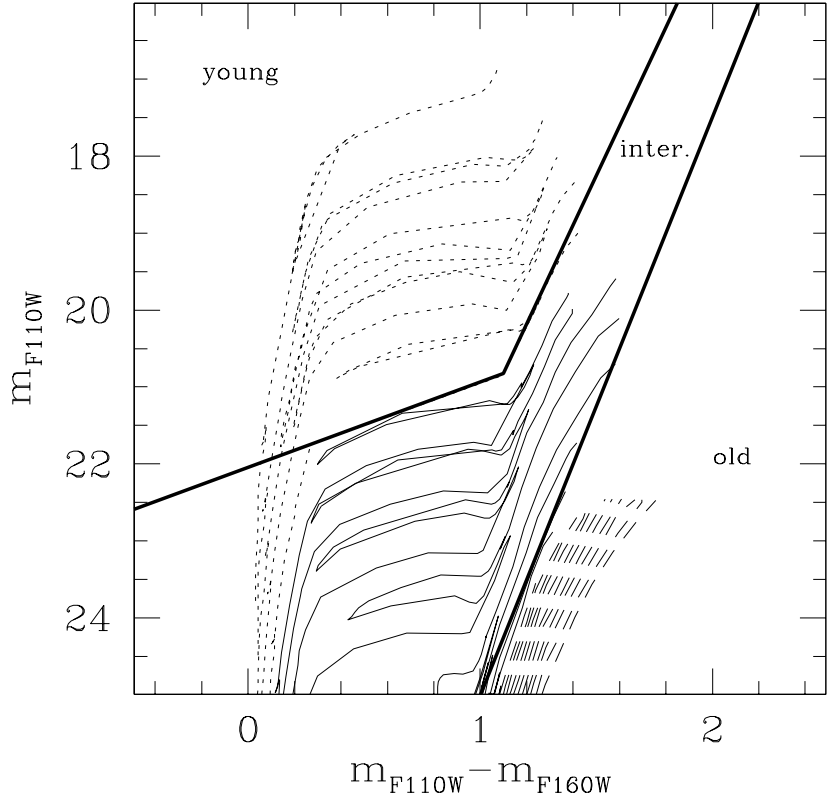}
\caption[fig]{Observational plane of the NIR CMD in F110W and F160W with Padua tracks 
at $Z=0.004$. The initial stellar masses from left to right are the following: 
30, 20, 15, 12, 9 \MSUN~(dotted lines), 7, 6, 5, 4, 3, 2.5, 2, 1.9 \MSUN~(solid lines), 
1.8, 1.7, 1.6, 1.5, 1.4, 1.3, 1.2, 1.1, 1.0, 0.9, 0.8 \MSUN~(dashed lines). The two 
straight lines along the 9 \MSUN~track divide the region of young stars ($\tau\,\lsim$50~ 
Myr) from that of intermediate age ones (\,50\,Myr$\lsim\tau\lsim$1\,Gyr). The other 
straight line along the 1.9 \MSUN~stellar track separates intermediate age from old 
stars ($\tau\,\gsim$1 Gyr). The resolved stars are omitted from the CMD for clarity.  
\label{tracks_distr}}
\end{figure}

\clearpage
\begin{figure}
\epsscale{0.9}
\plotone{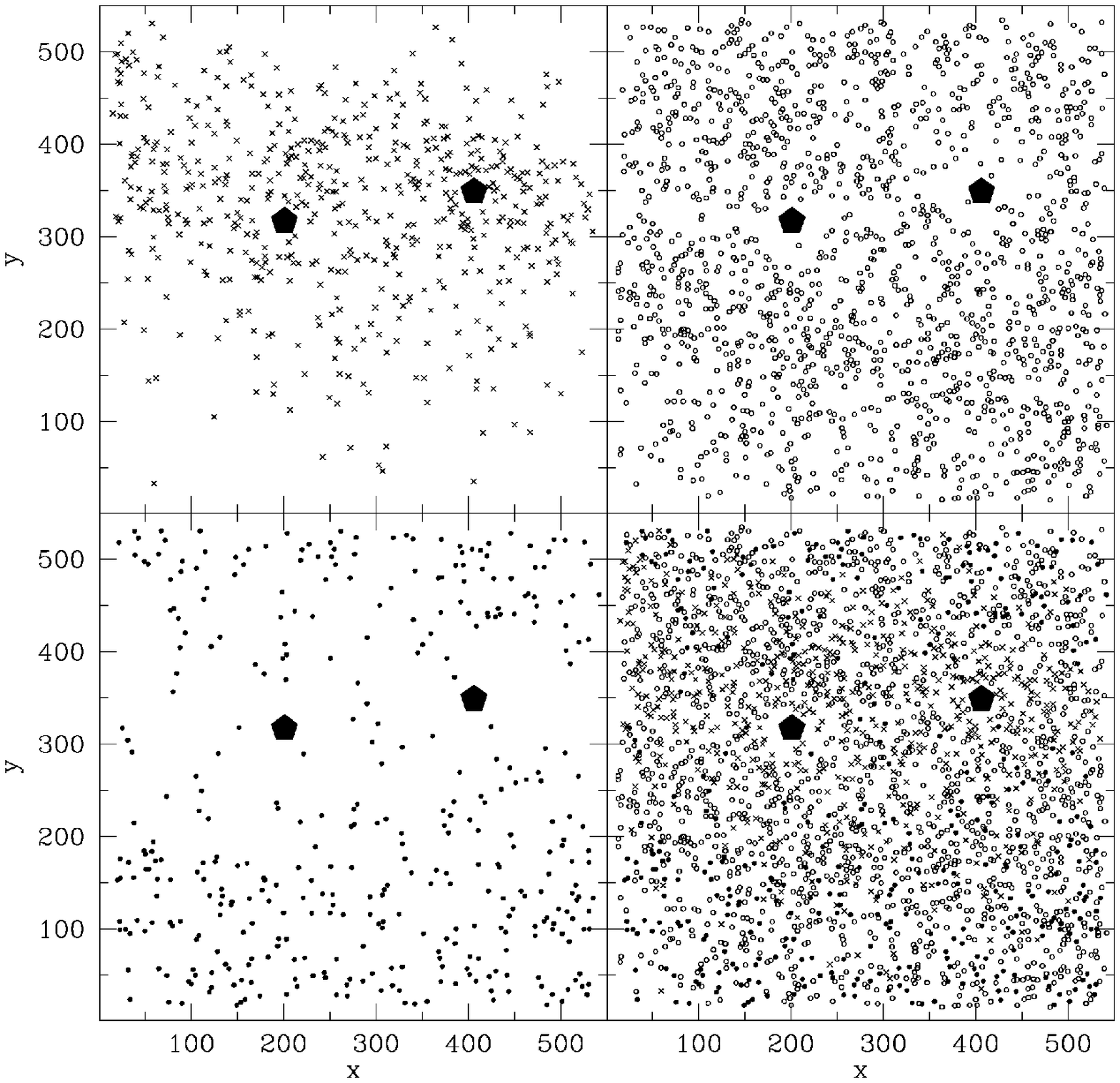}
\caption[fig]{Spatial distribution of different-age stars in the NIC2 field of view of
NGC~1569: young stars with ages less than 50 Myr (crosses in the top left panel), 
intermediate age stars from 50 Myr to 1 Gyr (open circles in the top right panel), old 
stars with ages more than 1 Gyr (filled circles in the bottom left panel). The bottom
right panel shows all the types of stars together. The two big filled symbols in each 
panel are the locations of the SSCs. The scale of the four panels is in pixels
(1 pixel = 0$\farcs$0375). Orientation is the same of Figure~\ref{imageH}.
\label{distr}}
\end{figure}


\clearpage

\begin{references}

\reference{} Aloisi, A., Tosi, M., \& Greggio, L. 1999, \aj, 118, 302

\reference{} Arp, H., \& Sandage, A. 1985, \aj, 90, 1163

\reference{} Babul, A., \& Ferguson, H.C. 1996, \apj, 458, 100

\reference{} Babul, A., \& Rees, M.J. 1992, \mnras, 255, 346

\reference{} Blitz, L., Spergel, D.N., Teuben, P.J., Hartmann, D., \&
Burton, W.B. 1999, \apj, 514, 818

\reference{} Bolte, M. 1989, \apj, 341, 168

\reference{} Buckalew, B.A., Dufour, R.J., Shopbell, P.L., \& Walter, D.K. 
2000, \aj, 120, 2402

\reference{} Calzetti, D. et al. 1999, ``NICMOS Instrument Handbook'', 
Version 3.0 (Baltimore: STScI)

\reference{} Calzetti, D., Kinney, A., \& Storchi-Bergmann, T. 1994,
\apj, 429, 582

\reference{} Casertano, S., Ratnatunga, K.U., \& Bahcall, J.N. 1990, 
\apj, 357, 435

\reference{} Cole, S., Aragon-Salamanca, A., Frenk, C. S., Navarro, J. F.,
\& Zepf, S. E. 1994, MNRAS, 271, 781

\reference{} Della Ceca, R., Griffiths, R.E., Heckman, T.M., \&
MacKenty, J.W. 1996, \apj, 469, 662

\reference{} De Marchi, G., Clampin, M., Greggio, L., Leitherer, C.,
Nota, A., \& Tosi, M.  1997, \apjl, 479, L27

\reference{} De Marchi, G., Nota, A., Leitherer, C., Ragazzoni, R.,
\& Barbieri, C. 1993, \apj, 419, 658

\reference{} D'Ercole, A., \& Brighenti, F. 1999, \mnras, 309, 941

\reference{} Devost, D., Roy, J.R., \& Drissen, L. 1997, \apj, 482, 765 

\reference{} Dickinson, M. 1999, ``NICMOS Data Handbook'', Version 4.0
(Baltimore: STScI)

\reference{} Diolaiti, E., Bendinelli, O., Bonaccini, D., Close, L.M., 
Currie, D.G., Parmeggiani, G. 2000, A\&AS, accepted, astro-ph/0009177

\reference{} Fagotto, F., Bressan, A., Bertelli, G., \& Chiosi,
C. 1994$a$, A\&AS, 104, 365

\reference{} Fagotto, F., Bressan, A., Bertelli, G., \& Chiosi,
C. 1994$b$, A\&AS, 105, 29

\reference{} Fanelli, M., O'Connell, R., \& Thuan, T. 1988, \apj, 334, 665

\reference{} Figer, D.F., Kim, S.S., Morris, M., Serabyn, E., Rich, R.M.,
\& McLean, I.S. 1999, \apj, 525, 750

\reference{} Fruchter, A.S., \& Hook, R.N. 1998, PASP, submitted, 
astro-ph/9808087

\reference{} Gallagher III, J.S., Hunter, D.A., \& Tutukov, A.V. 1984,
\apj, 284, 544

\reference{} Gonz\'alez-Delgado, R.M., Leitherer, C., Heckman, T.M.,
\& Cervi\~no, M. 1997, \apj, 483, 705

\reference{} Greggio, L., Tosi, M., Clampin, M., De Marchi, G.,
Leitherer, C., Nota, A., \& Sirianni, M. 1998, \apj, 504, 725 (G98)

\reference{} Greve, A., Becker, R., Johansson, L.E.B., \& McKeith, C.D.
1996, \aa, 312, 391

\reference{} Heckman, T.M., Dahlem, M., Lehnert, M.D., Fabbiano, G.,
Gilmore, D., \& Waller, W.H. 1995, ApJ, 448, 98

\reference{} Ho, L.C., \& Filippenko, A.V. 1996, \apjl, 466, L83

\reference{} Holfeltz, S.~T., \& Calzetti, D. 1999, ISR NICMOS-99-007

\reference{} Hunter, D.A., \& Gallagher III, J.S. 1986, \pasp, 98, 5

\reference{} Hunter, D.A., Hawley, W.N., \&  Gallagher III, J.S. 1993, \aj,
106, 1797

\reference{} Hunter, D.A., O'Connell, R.W., Gallagher III, J.S., \& 
Smecker-Hane, T.A. 2000, \aj, 120, 2383

\reference{} Israel, F.P. 1988, \aa, 194, 24

\reference{} Israel, F.P., \& van Driel, W. 1990, \aa, 236, 323

\reference{} Izotov, Y.I., \& Thuan, T.X. 1999, \apj, 511, 639

\reference{} Kauffmann, G., White, S.D.M., \& Guiderdoni, B. 1993, MNRAS, 264, 201

\reference{} Krist, J., \& Hook, R. 1999, ``The Tiny Tim User's Guide'',
Version 5.0 (Baltimore: STScI)

\reference{} Kunth, D. 1985, in Star Forming Dwarf Galaxies and Related 
Objects, ed. D. Kunth, T.X. Thuan, J. Tran Thanh Van (Edition Fronti\`eres, 
Paris, France), 183

\reference{} Legrand, F. 2000, A\&A, 354, 504

\reference{} Lilly, S.J., Tresse, L., Hammer, F., Crampton, D., \& 
Le Fevre, O. 1995, \apj, 455, 108

\reference{} Mac Low, M.M., \& Ferrara, A. 1999, \apj, 513, 142

\reference{} Melnick, J., Moles, M, \& Terlevich, R. 1985, A\&A, 149, L24

\reference{} Meurer, G. 1995, Nature, 375, 742

\reference{} M\"oller, C.S., Fritze-v.Alvensleben, U., \& Fricke, K.J. 1997,
\aa, 317, 676 

\reference{} O'Connell, R.W., Gallagher III, J.S., \& Hunter, D.A. 1994,
ApJ, 433, 65

\reference{} Origlia, L., \& Leitherer, C. 2000, \aj, 119, 2018

\reference{} Origlia, L., Leitherer, C., Aloisi, A., Greggio, L., \& Tosi, M. 
2000, submitted to AJ

\reference{} \"Ostlin, G. 2000, ApJ, 535, L99

\reference{} Papaderos, P., Loose, H.H., Fricke, K.J., \& Thuan, T.X.
1996$a$, \aa, 314, 59

\reference{} Papaderos, P., Loose, H.H., Thuan, T.X., \& Fricke, K.J.
1996$b$, A\&AS, 120, 207

\reference{} Prada, F., Greve, A., \& McKeith, C.D. 1994, \aa, 288, 396

\reference{} Roye, E.W., \& Hunter, D.A. 2000, \aj, 119, 1145

\reference{} Sandage, A. 1972, \apj, 178, 1

\reference{} Schulte-Ladbeck, R.E., Crone, M.M., \& Hopp, U. 1998, \apj
493, L23

\reference{} Schulte-Ladbeck, R.E., Hopp, U., Crone, M.M., \& Greggio, L.
2000$a$, in The First Stars, ed. Weiss, A., Abel, T.G., \& Hill, V. (Garching: 
ESO), in press

\reference{} Schulte-Ladbeck, R.E., Hopp, U., Greggio, L., \& Crone, M.M.
1999, \aj, 118, 2705

\reference{} Schulte-Ladbeck, R.E., Hopp, U., Greggio, L., \& Crone, M.M.
2000$b$, \aj, 120, 1713

\reference{} Searle, L., \& Sargent, W.L.W. 1972, \apj, 173, 25

\reference{} Searle, L., Sargent, W.L.W., \& Bagnuolo, 1973, \apj, 179, 427

\reference{} Stil, J.M., \& Israel, F.P. 1998, \aa, 337, 64

\reference{} Sweigart, A.V., Greggio, L., \& Renzini, A. 1990, \apj, 364, 
527

\reference{} Taylor, C.L., H\"uttemeister, S., Klein, U, \& Greve, A. 
1999, \aa, 349, 424

\reference{} Thuan, T.X. 1983, \apj, 268, 667

\reference{} Thuan, T.X., Izotov, Y.I., Lipovetsky, V., \& Pustilnik, 
S.A. 1994, in ESO/OHP Workshop on Dwarf Galaxies, ed. G. Meylan \& P. 
Prugniel (Garching: ESO), 421

\reference{} Thuan, T.X., \& Martin, G.E. 1981, \apj, 247, 823

\reference{} Tomita, A., Ohta, K., \& Saito, M. 1994, \pasj, 46, 335

\reference{} Tosi, M. 1999, in Dwarf galaxies and Cosmology, XXXIII
Rencontres de Moriond, T.X.Thuan, C.Balkowski, V.Cayatte, J.Tran
Thanh Van eds, (Edition Fronti\`eres, Paris, France), 443

\reference{} Tosi, M., Aloisi, A., Bellazzini, M., Clampin, M., Greggio, 
L., Leitherer, C., Nota, A., Origlia, L., \& Sabbi, E. 2000$a$, in preparation

\reference{} Tosi, M., Aloisi, A., Clampin, M., Greggio, L., Leitherer, C., 
Nota, A., \& Origlia, L. 2000$b$, in preparation

\reference{} Tosi, M., Greggio, L., Marconi, G., \& Focardi, P. 1991,
\aj, 102, 951 (TGMF)

\reference{} Tyson, A.J. 1988, \aj, 96, 1

\reference{} Vallenari, A., \& Bomans, D.J. 1996, \aa, 313, 713 (VB)

\reference{} Williams, R.E., Blacker, B., Dickinson, M. et al. 1996,
\aj, 112, 1335

\reference{} Young, J.S., Gallagher III, J.S., \& Hunter, D.A. 1984, \apj, 276, 476

\reference{} Zhang, Q., \& Fall, S.M. 1999, \apj, 527, L81

\end{references}
\end{document}